\documentclass{aa}  
\usepackage{graphicx}
\usepackage{txfonts}
\usepackage{natbib}
\bibpunct{(}{)}{;}{a}{}{,}   
\begin{document}
   \title{Bright globular clusters in NGC~5128: 
   the missing link between young massive clusters and evolved massive
   objects\thanks{Based 
on observations collected at the European Southern Observatory, Paranal,
Chile, within the Observing Programmes 63.N-0229 and 069.D-0169.
}
}

%
   
\author{M.~Rejkuba\inst{1}
	\and P.~Dubath\inst{2,3}
	\and D.~Minniti\inst{4}
	\and G.~Meylan\inst{5}
}

   \offprints{M. Rejkuba}

   \institute{ESO, Karl-Schwarzschild-Strasse
           2, D-85748 Garching, Germany\\
              \email{mrejkuba@eso.org}
	   \and
	      Observatoire de Geneve, ch.\ des Maillettes 51, CH-1290 Sauverny, 
	 Switzerland
	   \and
	      INTEGRAL Science Data Centre, ch.\ d'Ecogia 16, CH-1290 Versoix,
	 Switzerland\\
	      \email{pierre.dubath@obs.unige.ch}  
         \and
             Department of Astronomy and Astrophysics, Pontificia Universidad
	Cat\'olica de Chile, Vicu\~na Mackenna 4860, Santiago 22, Chile\\
             \email{dante@astro.puc.cl}
	  \and 
	     Ecole Polytechnique F\'ed\'erale de Lausanne (EPFL),
         Observatoire, CH-1290 Sauverny, Switzerland\\
	      \email{georges.meylan@epfl.ch} 
             }

   \date{Received 01 October 2006; accepted 13 March 2007}
\titlerunning{Bright globular clusters in NGC~5128}

 
  \abstract
   {Globular clusters are the simplest stellar systems in which 
   structural parameters are found to correlate with their masses and
   luminosities. 
   }
   {In order to investigate whether the brightest globular clusters in the giant
   elliptical galaxies are similar to the less luminous globular clusters like 
   those found in Local Group galaxies, we study  
   the velocity dispersion and structural parameter correlations of 
   a sample of bright globular clusters in the nearest giant elliptical
   galaxy NGC~5128 (Centaurus~A).}
   {UVES echelle spectrograph on the ESO Very Large Telescope (VLT)
   was used to obtain high resolution spectra of 23 bright globular 
   clusters in NGC~5128, and 10 clusters were observed with EMMI in
   echelle mode with the ESO New Technology Telescope. The two datasets have 5
   clusters in common, while one cluster observed with UVES had too low
   signal-to-noise (S/N). Hence the total number of clusters analysed in this work 
   is 27, more than doubling the previously 
   known sample. Their spectra were cross-correlated with template spectra 
   to measure the central velocity dispersion for each target. The structural 
   parameters were either taken from the existing literature, or in cases where this
   was not available, we have derived them from our VLT FORS1 images taken under
   excellent seeing conditions, using the ISHAPE software. The velocity
   dispersion and structural parameter measurements were used to obtain masses and
   mass-to-luminosity  ratios ($M/L_V$) of 22 clusters.}
   {The masses of the clusters in our sample range from 
   $M_{vir}=10^5-10^7$~M\sun~ and the average $M/L_V$ is $3 \pm 1$.
   The three globular clusters harbouring X-ray point sources are 
   the second, third and sixth most massive in our sample. 
   The most massive cluster, HCH99-18, is also the brightest and the largest 
   in size. It has the mass ($M_{vir}=1.4\times 10^7$~M\sun) an order of 
   magnitude larger than the most massive clusters in the Local Group, and  
   a high $M/L_V$ ratio ($4.7 \pm 1.2$). We discuss briefly possible
   formation scenarios for this object.}
   {The correlations of structural 
   parameters, velocity dispersion, masses and $M/L_V$ for the bright globular
   clusters in NGC~5128 extend the properties established for the 
   most massive Local Group clusters towards 
   those characteristic of dwarf elliptical galaxy nuclei and Ultra 
   Compact Dwarfs (UCDs). The detection of the mass--radius and the 
   mass--$M/L_V$ relations for the globular clusters with masses greater than 
   $\sim 2 \times 10^6$~M\sun~ provides the missing link between ``normal'' old
   globular clusters, young massive clusters, and evolved objects like UCDs.}


   \maketitle
%

\section{Introduction}
\label{sect:intro}

The properties of globular clusters and the observed correlations between 
their various internal structural and dynamical parameters offer 
empirical  constraints not only on the formation of globular clusters 
themselves, but also on the history of the host galaxy. A large number
of  empirical  relations between various properties, core and half-light 
radii, surface brightnesses, velocity dispersions, concentrations, 
luminosities, metallicities, etc., of the Milky Way globular clusters 
have been found \citep[e.g.][]{djorgovski+meylan94}. 
Many of them are mutually dependent due to the fact that globular 
clusters have remarkably simple structures that can be reasonably well 
approximated by isotropic, single-mass  \citet{king66} models. 
\citet{mclaughlin00} has shown that the Milky Way 
clusters are confined to the fundamental plane well defined by 2 empirical
relations: M/L=const. and E$_b\sim L^{2.05}$ ,where M is the mass, L the 
luminosity, and E$_b$ the binding energy of the cluster. 

The  brightest and the most  massive globular cluster
of our  Galaxy, $\omega$ Cen \citep{meylan+95}, is
peculiar in many of its characteristics: e.g. it is the most flattened
Galactic globular cluster  \citep{white+shawl87} and it
shows strong variations in  nearly all element abundances
\citep[e.g.][]{norris+dacosta95,pancino+02}.  
A  scenario that may explain
some of its  characteristics is that $\omega$ Cen is  the nucleus of a
former dwarf  elliptical galaxy 
\citep{zinnecker+88,hughes+wallerstein00,hilker+richtler00}.
The  same scenario  was
proposed  for M54,  another very  massive globular  cluster.  It  is a
candidate for the  nucleus of the Sagittarius dwarf 
\citep[e.g.][]{bassino+muzzio95,layden+sarajedini00}, 
a  galaxy  that is  currently  being
accreted  by the Milky  Way.  M31  has 4  globular clusters  for which
\citet{djorgovski+97} measured  velocity dispersions
$\sigma>20$~km~s$^{-1}$ implying  masses at least as large  as the one
of  $\omega$ Cen.  The  most massive  of them,  
G1, shares  also other particular  properties  of  $\omega$  Cen,  
like  the  flattening  and metallicity dispersion \citep{meylan+01}. 
From the recent work of \citet{ma+06}, the most luminous M31 globular 
cluster, 037-B327,  has been suggested to be the most massive
Local Group cluster, with a total mass of $(3.0 \pm 0.5)\times 10^7$~M\sun,
determined  photometrically. These authors estimate the 
(one-dimensional) velocity
dispersion for 037-B327 of $(72 \pm 13$~km~s$^{-1}$).  However, a later
paper by \citet{cohen06} challenged this result, based on the measured velocity 
dispersion of $\sigma=21.3 \pm 0.4$~km~s$^{-1}$, which is comparable to that of G1
\citep[$\sigma=25.1 \pm 0.3$~km~s$^{-1}$;][]{djorgovski+97}. 
She concluded that 037-B327 is not the most massive cluster in the Local Group,
and that probably M31 clusters G1, G78 and G280 are more massive than 037-B327. 
Going to galaxies beyond the Local Group, 
very similar to G1 in M31 is the cluster n1023-13 in NGC~1023 \citep{larsen01}. 

With luminosities and masses larger than globular clusters are the so-called
ultra compact dwarfs (UCDs) or dwarf-globular transition objects (DGTOs) 
discovered in Fornax and Virgo galaxy clusters 
\citep{hilker+99,drinkwater+00,hasegan+05,hilker+07}. 
While their origin and relation
to globular clusters is still debated in the literature, it has been established that
very massive young star clusters can form in major star forming events. Such clusters,
with masses of the order of $>10^6$, or even $>10^7$~M\sun~ \citep{maraston+04,
bastian+06}, show similar scaling relations as UCDs/DGTOs, but might be different from
the less massive globular clusters based on examination of
their mass-velocity dispersion and mass-radius relations \citep{kissler-patig+06}.
However, we point out that the ages of UCDs/DGTOs are similar to those of globular
clusters, in contrast to the massive young star clusters forming in mergers.

To populate the transition region between ``normal'' globular clusters 
and more massive DGTOs, it is of interest to look at the
massive elliptical galaxies which harbour globular cluster systems which 
are an order of magnitude more populous than those of the Local Group 
spiral galaxies. The nearest easily observable elliptical galaxy is NGC~5128. 
It has a large number of bright globular clusters with luminosities exceeding 
the brightest Local Group globulars. This makes it an ideal target. 

The most recent distance determination to this
galaxy is $3.42 \pm 0.18 \pm 0.25$~Mpc (the first is random and 
the second systematic error), obtained 
using Cepheid PL relation \citep{ferrarese+06}. Here we use the distance of
$3.84 \pm 0.35$~Mpc \citep{rejkuba04}, which is the same value 
used in a previous work by \citet{martini+ho04} who presented velocity 
dispersions and mass-to-light ratios for 
14 bright globular clusters in NGC~5128. In this work we present new 
high resolution spectra and derive M/L ratios, thus more
than doubling the sample of bright globular clusters with similar data 
in the literature.

A decade ago \citet{dubath94} presented at a conference the first measurements 
of velocity dispersions of 10 bright globular clusters in NGC~5128.
Since these results have not been published
in a refereed journal yet, they are included here along with the more recent 
observations of 23 clusters from UVES high-resolution echelle spectrograph 
of ESO Kueyen (UT2) telescope of Very Large Telescope (VLT). 

This paper is organized as follows: Sect.~\ref{sect:data} describes the 
observations and data reduction, Sect.~\ref{sect:templatesRV} shows the 
results of the cross-correlation technique for radial velocity and 
metallicity standard stars, while Sect.~\ref{sect:gcRV} presents the 
results from the radial velocity and core velocity dispersion measurements
of globular clusters in NGC~5128. The comparison with previous 
measurements of clusters' radial velocity and velocity dispersion is 
in Sect.~\ref{sect:compare}. In Sect.~\ref{sect:structure} 
the structural parameters for 22 clusters are presented. 
For those clusters which had no 
previous determinations of structural parameters in the literature we derive 
them from our high resolution ground-based images 
fitting the King profile \citep{king62} using ISHAPE \citep{larsen99,larsen01} 
programme. 
In Sect.~\ref{sect:ML} we derive mass-to-luminosity ratios for the 
clusters and in Sect.~\ref{sect:FP} 
discuss the correlations and fundamental plane. Finally, in 
Sect.~\ref{sect:summary} we summarize our results.


\section{Sample selection and observations}
\label{sect:data}

The observations of 10 bright clusters 
\citep[selected from the lists of][]{VHH81, hesser+86, harris+92}
were taken in March-April 1993 with the echelle mode of EMMI \citep{emmi}, 
the multi-mode
instrument of the ESO New Technology Telescope (NTT). These data have previously
been presented at a conference \citep{dubath94}, 
and are published here together with the
observations of 23 clusters obtained with UVES echelle spectrograph \citep{uves} 
of the ESO VLT in April 2002. 
There are 5 clusters observed with both instruments and these have
been used to check for the systematics in the data and errors.

The sample of globular clusters selected for 
observations with UVES contains the brightest NGC~5128 clusters with either membership 
confirmed through published radial velocities \citep[][cluster names 
starting with VHH81, HHH86, and HGHH92]{VHH81, hesser+86, harris+92}
or the structural parameters and colours typical for globular clusters 
in the Milky Way 
\citep[][cluster names starting with HCH99 and with R, respectively]{holland+99,
rejkuba01}. 

\subsection{EMMI spectroscopy}
\label{sect:EMMI-data}

The first high-resolution integrated-light spectra of bright globular clusters 
in NGC~5128 were obtained with EMMI at ESO NTT telescope 
during three nights, March $31$ to April $2$ 1993. 
The red arm of EMMI was used in 
Echelle mode (REMD) with grating \#10 and grism \#3 (CD2), yielding the 
resolving power of 30,000, corresponding to 10~km~s$^{-1}$, and the wavelength 
coverage was from 4500 to 9000 \AA, divided among 65 useful orders. 

In total 14 spectra of 10 of the brightest globular clusters, selected 
from the catalogues of \citet{VHH81}, \citet{hesser+84}, and 
\citet{harris+92}, were secured. The ThAr calibration lamp spectra were taken 
before and after each cluster spectrum. In addition, the following four K giant 
radial velocity standard stars were observed on each of the three nights: 
NGC~2447-s28, NGC~2447-s4, HD~171391 and HD~176047.
All the spectra were reduced with the INTER-TACOS software developed by Queloz 
\& Weber in Geneva Observatory \citep[see e.g.][]{queloz+95}.

\subsection{UVES spectroscopy}
\label{sect:UVES-data}

The UVES observations were carried out on the nights of 19 and 
20 April 2002 in visitor mode. 
The red arm of UVES spectrograph was used with the standard CD\#3 setting 
centered on 580~nm. It is equipped with two CCDs,  
covering the total wavelength range from 4760 \AA\, to 6840 \AA, with a gap 
of 50 \AA\, centered on 5800 \AA. The slit was 1" wide, giving the resolution 
of $\lambda / \Delta \lambda \sim 42,000$. 
The sky conditions 
were clear and the seeing varied between $0\farcs 6$ to $1 \farcs 3$, but it 
stayed most of the time around $0\farcs 8$. 

Globular clusters observed with UVES have $V$-band magnitudes ranging from 
17.1 to 18.8 for 22 clusters. The faintest observed cluster had $V=19.44$. 
The typical exposure times were 1200 sec for the brighter or 1800 sec for 
the fainter clusters, except for the faintest 19.4 mag cluster which was 
exposed for 2700 sec. Four clusters have been observed twice and one cluster 
three times during the two night run. The
multiple exposures have been averaged to increase the signal-to-noise (S/N), 
but were also reduced independently in order to provide estimates of 
measurement errors. The observation log for all the clusters is in 
Table~\ref{tab:obslog_gcs}, where we list (1) the name, (2) the observation date, (3)
the exposure times in seconds, (4) the typical S/N measured on the blue side 
of the $H_\alpha$ line at $\sim 6550$~\AA\, using the {\it splot} IRAF task. 
The last column lists the $V$ magnitude of the
clusters taken from the literature.

\begin{table}
\caption[]{Observations log for globular clusters: 
observations in 1993 
were done with EMMI at NTT and
in 2002 with UVES at Kueyen VLT. The nomenclature of the clusters is following
that of the \citet{peng+04GCcat} catalogue, and the  
magnitudes given in the
last column are from the same catalogue where available. For the clusters 
for which there are no measurements in that catalogue, 
we take the magnitudes from the original discovery publications.
}
\label{tab:obslog_gcs}
\begin{tabular}{llclr}
\hline \hline
\multicolumn{1}{c}{(1)} & \multicolumn{1}{c}{(2)} & \multicolumn{1}{c}{(3)}  & 
\multicolumn{1}{l}{(4)}&
\multicolumn{1}{c}{(5)}  \\
\multicolumn{1}{c}{ID} & \multicolumn{1}{c}{Date}  & Exp. & \multicolumn{1}{l}{S/N@}    
& \multicolumn{1}{c}{V} \\  
\multicolumn{1}{c}{} & \multicolumn{1}{c}{yyyy-mm-dd}  & sec &
\multicolumn{1}{l}{6550\AA}    
& \multicolumn{1}{c}{mag} \\  
\hline
      HGHH92-C1 & 1993-04-02& 4800 &       &  17.42  \\
      HGHH92-C1 & 2002-04-20& 1200 &  4    &  17.42  \\
      HGHH92-C1 & 2002-04-20& 1200 &  4.5  &  17.42  \\
       VHH81-C3 & 1993-04-01& 4800 &       &  17.71  \\
       VHH81-C5 & 1993-04-02& 4800 &       &  17.68  \\
      HGHH92-C6 & 1993-03-31& 3600 &       &  17.21  \\
      HGHH92-C6 & 1993-04-01& 4200 &       &  17.21  \\
      HGHH92-C7 & 1993-03-31& 3060 &       &  17.17  \\
      HGHH92-C7 & 1993-03-31& 3600 &       &  17.17  \\
      HGHH92-C7 & 2002-04-19& 1200 &  10   &  17.17  \\
      HGHH92-C7 & 2002-04-19& 1200 &  11   &  17.17  \\
      HGHH92-C7 & 2002-04-20& 1200 &  11   &  17.17  \\     
      HGHH92-C11& 2002-04-20& 1200 &  6    &  17.91  \\
      HGHH92-C11& 2002-04-20& 1200 &  7    &  17.91  \\
      HGHH92-C12=R281 & 1993-04-01& 4200 &	 &  17.74 \\
      HGHH92-C12=R281 & 2002-04-19& 1800 &  6    &  17.74 \\
       HHH86-C15=R226 & 2002-04-19& 1800 &  5	 &  18.56 \\
      HGHH92-C17& 1993-04-01& 4800 &       &  17.63  \\
      HGHH92-C17& 1993-04-02& 4500 &       &  17.63  \\
       HHH86-C18& 1993-04-02& 4500 &       &  17.53  \\
     HGHH92-C21 & 1993-04-01& 4200 &       &  17.87  \\
     HGHH92-C21 & 2002-04-20& 1200 &  5.5  &  17.87  \\
     HGHH92-C22 & 2002-04-20& 1800 &  7.5  &  18.15  \\
     HGHH92-C23 & 1993-03-31& 4500 &       &  17.22  \\
     HGHH92-C23 & 1993-04-02& 4500 &       &  17.22  \\
     HGHH92-C23 & 2002-04-19& 1200 &  10   &  17.22  \\
     HGHH92-C23 & 2002-04-20& 1200 &   9   &  17.22  \\
     HGHH92-C29 & 2002-04-20& 1200 &  6.5  &  18.15  \\
     HGHH92-C36=R113 & 2002-04-19& 1800 &  5.5	&  18.35 \\
     HGHH92-C37=R116 & 2002-04-19& 1800 &  6.5	&  18.43 \\
      HHH86-C38=R123 & 2002-04-20& 1800 &  6	&  18.41 \\
     HGHH92-C41 & 2002-04-20& 1800 &  6    &  18.59  \\
     HGHH92-C44 & 2002-04-19& 1800 &  4    &  18.69  \\
     HGHH92-C44 & 2002-04-19& 1800 &  5    &  18.69  \\
       HCH99-2  & 2002-04-20& 1200 &  4    &  18.21  \\
       HCH99-15 & 2002-04-19& 1200 &  6    &  17.56  \\
       HCH99-16 & 2002-04-20& 1800 &  3.5  &  18.45  \\
       HCH99-18 & 2002-04-19& 1200 &  8    &  17.07  \\
       HCH99-21 & 2002-04-20& 1208 &  3.5  &  18.41  \\
        R115    & 2002-04-20& 2700 &  2    &  19.44  \\
        R122    & 2002-04-19& 1800 &  7    &  18.09  \\
        R223    & 2002-04-19& 1800 &  5    &  18.77  \\
        R261    & 2002-04-19& 1800 &  5.5  &  18.20  \\
\hline
\end{tabular}
\end{table}

Apart from the globular cluster targets, we have observed 17 different G and
K-type giant stars with a range of metallicities ($-2.6 <\mathrm{[Fe/H]}<
+0.3$~dex) to be used as templates for
cross-correlation. Some stars were observed several times, thus yielding 
a total of  28 high S/N stellar spectra.
The observation log of the template stars is in Table~\ref{tab:obslog_stars}. 

\begin{table*}
\caption[]{Template stars observed with UVES during the 2002 run. 
The columns list: (1) identifier, (2) number of observed spectra, (3) spectral type, 
(4) apparent V band magnitude 
from the literature, (5) metallicity from the literature, (6) radial velocity from the literature,
(7) measured radial velocity, (8) reference for catalogue value of radial velocity,
and average $\sigma_{CCF}$ measured from cross-correlation with all the other stars
for (9) lower CCD, and (10) upper CCD (see Eq.~\ref{eq:sigma_ccf} for the definition of $\sigma_{CCF}$).}
\label{tab:obslog_stars}
\begin{tabular}{lclrrrrlrr}
\hline \hline
\multicolumn{1}{c}{(1)} & \multicolumn{1}{c}{(2)} & \multicolumn{1}{c}{(3)}  & \multicolumn{1}{c}{(4)}&
\multicolumn{1}{c}{(5)} & \multicolumn{1}{c}{(6)} &\multicolumn{1}{c}{(7)} & \multicolumn{1}{c}{(8)}&
\multicolumn{1}{c}{(9)} & \multicolumn{1}{c}{(10)}\\
\multicolumn{1}{c}{ID} & N & Sp.\ Typ  & V (mag)   & \multicolumn{1}{c}{[Fe/H]} 
& V$_R$(cat) &V$_R$(UVES) &\multicolumn{1}{l}{Ref.}&
\multicolumn{1}{c}{$\sigma_{CCF}^l$} &\multicolumn{1}{c}{$\sigma_{CCF}^u$}\\  
\hline
HD~103295 & 2 &G5/G6~III &9.60  &$-1.01$ &$  3.0 \pm 0.3$ &$ -2.6 \pm 0.3$  &N04 &$12.2 \pm 0.3$&$11.5 \pm 0.2$ \\
HD~107328 & 2 &K1~III    &5.00  &$-0.48$ &$ 36.40\pm0.01$ &$ 36.4 \pm 0.1$  &F05 &$12.4 \pm 0.2$&$11.7 \pm 0.1$ \\
HD~146051 & 3 &M1~III    &2.74  &$+0.32$ &$-19.6 \pm 0.3$ &$-20.1 \pm 0.3$  &COR &$12.9 \pm 0.2$&$12.0 \pm 0.1$ \\
HD~150798 & 1 &K2~II-III &1.92  &$-0.06$ &$ -3.0 \pm 0.3$ &$ -3.0 \pm 0.1$  &COR &$13.0 \pm 0.2$&$12.2 \pm 0.1$ \\
HD~161096 & 1 &K2~III    &2.77  &        &$-12.53\pm0.01$ &$-12.5 \pm 0.1$  &F05 &$12.6 \pm 0.2$&$11.8 \pm 0.1$ \\
HD~165195 & 1 &K3~III    &7.34  &$-2.24$ &$ -0.3 \pm 0.2$ &$ -0.5 \pm 0.2$  &F05 &$12.7 \pm 0.5$&$11.9 \pm 0.2$ \\
HD~168454 & 1 &K3~III    &2.71  &        &$-20.4 \pm 0.3$ &$-20.5 \pm 0.1$  &COR &$12.7 \pm 0.2$&$11.9 \pm 0.1$ \\
HD~171391 & 1 &G8~III    &5.13  &$-0.07$ &$  7.4 \pm 0.2$ &$  7.3 \pm 0.2$  &COR &$12.5 \pm 0.2$&$11.7 \pm 0.1$ \\
HD~196983 & 2 &K2~III    &9.08  &        &$ -9.1 \pm 0.3$ &$ -9.1 \pm 0.2$  &COR &$12.6 \pm 0.2$&$11.7 \pm 0.1$ \\
HD~203638 & 3 &K0~III    &5.37  &$+0.30$ &$ 22.1 \pm 0.2$ &$ 22.1 \pm 0.1$  &COR &$12.6 \pm 0.2$&$11.7 \pm 0.1$ \\
HD~33771  & 2 &G0~III    &9.50  &$-2.56$ &$-13.6 \pm 0.4$ &$-13.6 \pm 0.2$  &D97 &$12.4 \pm 0.5$&$11.7 \pm 0.3$ \\
HD~66141  & 3 &K2~III    &4.40  &$-0.36$ &$ 71.57\pm0.01$ &$ 71.5 \pm 0.4$  &F05 &$12.4 \pm 0.2$&$11.6 \pm 0.1$ \\
HD~81797  & 2 &K3~III    &1.99  &$-0.12$ &$ -4.7 \pm 0.3$ &$ -4.7 \pm 0.3$  &COR &$12.7 \pm 0.2$&$11.9 \pm 0.1$ \\
HD~83212  & 1 &G8~IIIw   &8.34  &$-1.51$ &$108.7 \pm 0.3$ &$109.1 \pm 0.1$  &D97 &$12.4 \pm 0.3$&$11.7 \pm 0.2$ \\
HD~93529  & 1 &G6/G8w    &9.31  &$-1.56$ &$145.4 \pm 0.3$ &$144.9 \pm 0.3$  &D97 &$12.2 \pm 0.4$&$11.5 \pm 0.2$ \\
NGC~2447-s28&1&G8/K0~III &10.15 &$+0.10$ &$ 21.2 \pm 0.1$ &$ 21.8 \pm 0.1$  &D97 &$12.4 \pm 0.2$&$11.7 \pm 0.1$ \\
NGC~2447-s4 &1&G8/K0~III &9.85  &$+0.10$ &$ 23.2 \pm 0.2$ &$ 23.1 \pm 0.2$  &D97 &$12.4 \pm 0.2$&$11.7 \pm 0.1$ \\
\hline
\multicolumn{8}{l}{References: All the radial velocities except those 
from D97 \citep{dubath+97} are compiled from}\\
\multicolumn{8}{l}{http://www.casleo.gov.ar/catalogue/catalogue.html, which lists 
references to the sources:}\\ 
\multicolumn{8}{l}{N04 \citep{nordstrom+04}; F05 \citep{famaey+05}; }\\ 
\multicolumn{8}{l}{COR \citep[][see also:
http://obswww.unige.ch/~udry/std/stdcor.dat]{udry+99}.}
\end{tabular}
\end{table*}

After each target spectrum, globular cluster or star, we have obtained the 
ThAr lamp spectrum at the same telescope position. The bias and flat-field 
calibration data were taken at the end of each night. 

The data reduction was done both using the {\it echelle} package in IRAF 
\citep{echelleman} and the MIDAS based ESO-UVES pipeline \citep{ballester+00}, 
where we have taken care to assign the wavelength calibration spectrum taken 
after each target spectrum in order to have the highest precision in the 
wavelength calibration. 
Due to low S/N ratio of the spectra, the MIDAS pipeline was not used in optimal 
extraction mode. The final spectra were normalized using the {\it continuum} task 
in IRAF and the cosmic rays were excised by hand. After some tests to ensure 
that the pipeline results were giving the same results as manual reductions done 
within IRAF, we have decided to later use spectra reduced within the MIDAS 
pipeline because the different echelle orders were combined in one long 1D 
spectrum per CCD, offering thus the maximum number of lines for cross-correlation.


\section{Cross-correlation}

In order to measure radial velocities and velocity dispersions of all our 
targets we have used cross-correlation technique \citep{tonry+davis79}. Slightly
different implementation of the cross-correlation technique has been adopted for 
the EMMI and UVES spectra.  The comparison of the resulting velocity dispersion 
measurements for 
the 5 targets in common between the two datasets is providing a useful 
check on the results obtained with these two slightly different methods.

EMMI spectra were cross-correlated with a numerical mask specially designed 
for globular clusters. This has been described in greater detail
by \citet{dubath+90, dubath+92} and thus we do not repeat it here.

The globular cluster spectra observed with UVES have been cross-correlated with 
the high signal to noise spectra of template radial velocity stars 
observed during the same run, using the IRAF task {\it FXCOR} in the {\it RV} package.
All the spectra were
Fourier-filtered prior to cross-correlation, to remove the residual low frequency 
features arising from imperfect continuum fitting to the spectra with combined 
echelle orders. The features at frequencies higher than the intrinsic resolution of 
the spectrograph were also cut. The peak of the cross-correlation function (CCF)
traces the radial velocity, and the width is a function of the velocity 
dispersion and of the instrumental width. The latter is measured by 
cross-correlating the template stars spectra with each other. 
Since we have observed a large number of radial velocity standard stars as well 
as giant star templates with a range of metallicities, we could check that the 
template miss-match does not produce spurious results. This is described in  
detail in the next section.

\subsection{Template stars}
\label{sect:templatesRV}

The measured radial velocities for all the stars observed during the 2002 
run with UVES are given in Table~\ref{tab:obslog_stars} in column 7. 
They were measured by cross-correlating each Fourier-filtered stellar spectrum
against each other. The resulting radial velocities for each individual 
spectrum were averaged and we report here the average radial velocity 
and standard deviation for each star. In column 6 we list the radial velocity 
from the literature. These were compiled from
\citet{dubath+97} and the web database of stellar radial velocities 
(see table footnote). In all but one case the difference between our 
measured radial velocities from UVES spectra and those from the literature 
is smaller than 1~km~s$^{-1}$ and the measurements are 
consistent with the catalogue values within the errors. Since the stars were 
observed with a 1.0 arcsec slit and seeing was sometimes as good as 0.6-0.7 
arcsec, part of the error in radial velocities may come from slit centering 
errors. The star that shows the largest difference, HD~103295, has less certain 
value of the radial velocity and the quoted error from the literature is 
evidently underestimated. Leaving HD~103295 out, the average difference between 
our radial velocity measurements and catalogued values is 
$V_R(UVES)-V_R(cat) = 0.07 \pm 0.27$~km~s$^{-1}$ indicating that the systematic 
errors due to slit centering are not significant.  For the cross-correlation 
of stars with cluster spectra we adopt our measured radial velocity 
for HD~103295, and literature values for all the other stars.

\begin{figure}
\centering
\resizebox{\hsize}{!}{
\includegraphics[angle=270]{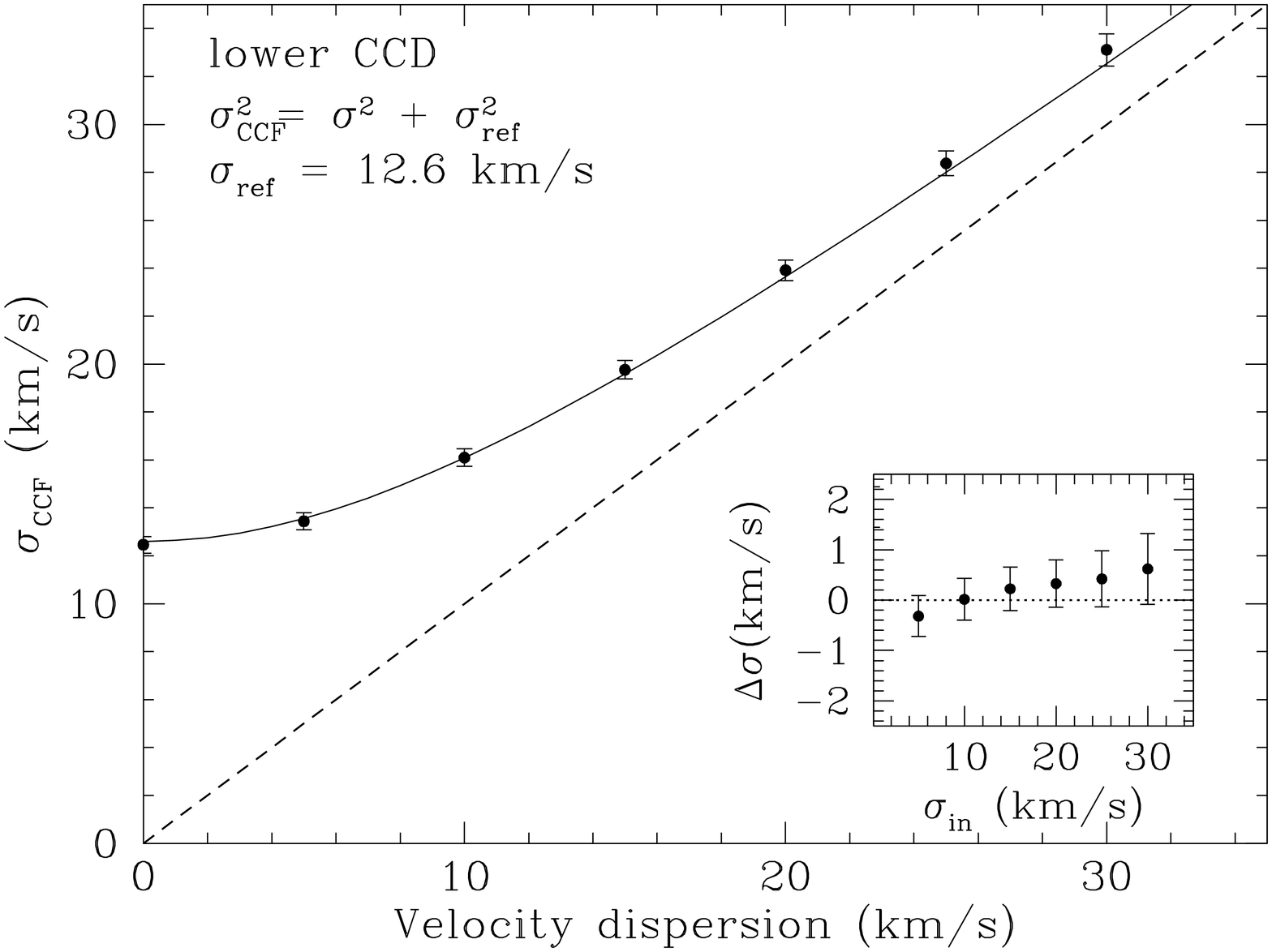}}
\resizebox{\hsize}{!}{
\includegraphics[angle=270]{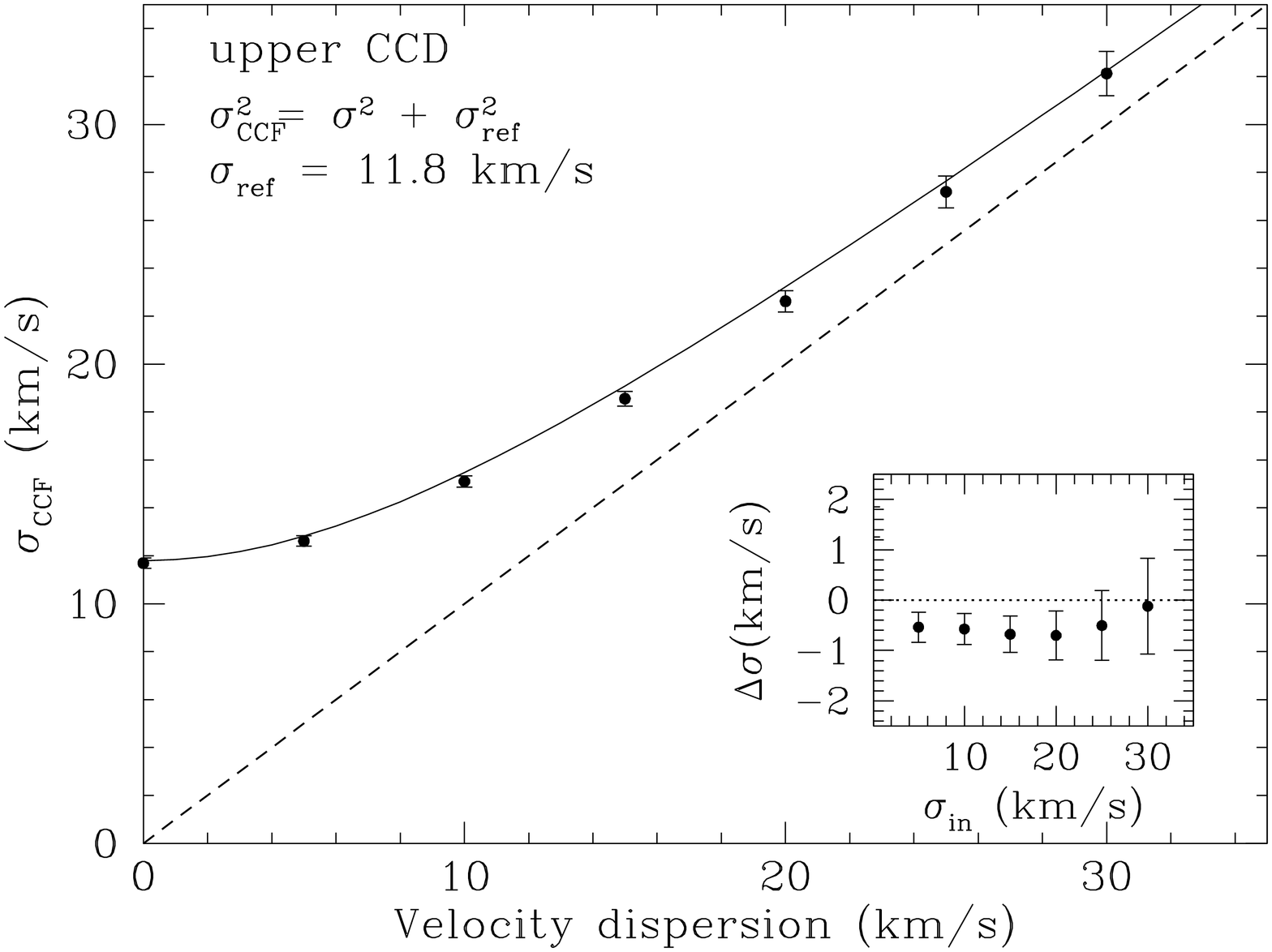}
}
\caption[]{The points with error bars are the
average raw velocity dispersion measurements ($\sigma_{CCF}$)
obtained from 7 different template stars whose spectra were broadened to simulate the
different input velocity dispersion ($\sigma_{in}$ values). The solid line is from 
Equ.~\ref{eq:sigma_ccf} which is also displayed in each panel. 
The two diagrams are for the
lower and upper CCDs of UVES which have different $\sigma_{ref}$ values. The inserts show
the difference between the obtained and input true velocity dispersion as a function of
input velocity dispersion.
}
\label{fig:sigmaw_ccf}
\end{figure}

The projected velocity dispersions ($\sigma$) for globular clusters 
are derived from the 
broadening of the cluster cross-correlation function (CCF) produced by the 
Doppler line broadening present in the integrated-light spectra due to the 
random spatial motion of stars. The raw measurement ($\sigma_{CCF}$) is 
however a quadratic sum of the $\sigma$ and the intrinsic 
instrumental width ($\sigma_{ref}$) \citep{dubath+92}:
\begin{equation}
\sigma_{CCF}^2 = \sigma^2 + \sigma_{ref}^2
\label{eq:sigma_ccf}
\end{equation}
The $\sigma_{ref}$ value is determined for both UVES CCDs by taking 
the average value of $\sigma_{CCF}$ measurements obtained by cross-correlating 
18 selected best template stellar spectra, belonging to 13 different stars, 
with all the other stars. This same set of stars is used in cross-correlation 
of globular cluster spectra as well as in all the simulations (see below). 
Since all these stars are late-type giants they 
are not expected to exhibit line broadening due to rotation. For 3 stars, 
HD~66141, HD~107328, and HD~161096, \citet{demedeiros+mayor99} provide 
measurements of rotational velocities which are, $1.1$, 1.3, and 
$<1.0$ km~s$^{-1}$, respectively, with uncertainties which are of the same 
order of magnitude. The fact that for all the stars CCF has similar width 
(see Table~\ref{tab:obslog_stars})
implies that rotation is not a concern. The weighted average of the stellar CCFs
are $12.6 \pm 0.2$~km~s$^{-1}$ for the lower
and $11.8 \pm 0.2$~km~s$^{-1}$ for the upper CCD. The difference in the 
instrumental width reflects the different resolution of the two spectral ranges.
Excluding from the average HD~150798, the star that shows the widest 
cross-correlation peak for both spectral ranges, does not change the 
average value of $\sigma_{ref}$.

Figure~\ref{fig:sigmaw_ccf} shows the validity of the Equ.~\ref{eq:sigma_ccf} 
for the lower and upper CCDs of UVES. The solid line is from Equ.~\ref{eq:sigma_ccf} 
and the points represent the average projected velocity dispersion measurements 
obtained from 7 different template stars whose spectra were broadened by 
convolving each of them with Gaussians with known sigma ($\sigma_{in}$) of 5, 10, 15, 
20, 25 and 30~km~s$^{-1}$. The velocity dispersion was then measured 
on these broadened spectra by convolving them with the 18 selected best 
template stellar spectra and averaging the resulting velocity dispersions. The smaller
inserts in each of the panels in Fig.~\ref{fig:sigmaw_ccf} show the difference between
the average measured and input velocity dispersion as a function of the input velocity
dispersion, while the main panel displays the averages of the raw measurements
($\sigma_{CCF}$) at each $\sigma_{in}$ value.


\begin{figure}
\centering
\resizebox{\hsize}{!}{
\includegraphics[angle=270]{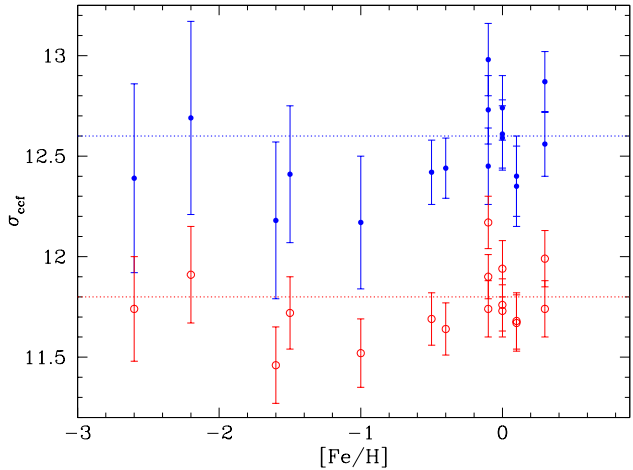}
}
\caption[]{Width of the cross-correlation function for stars is plotted as a 
function of metallicity. The measurements on the lower CCD are plotted with 
filled and on the upper CCD with open circles.
}
\label{fig:feh_ccf}
\end{figure}

We have carefully selected the regions for cross-correlation, avoiding strong
lines such as $H_{\beta}, H_{\alpha}$, sodium region, as well as Mgb region. When
we included for example $H_{\alpha}$ line, we noticed a strong trend of
cross-correlation width as a function of metallicity. In the final selection,
this dependence is not present as can be seen from Fig.~\ref{fig:feh_ccf}.

\begin{figure}
\centering
\resizebox{\hsize}{!}{
\includegraphics[angle=270]{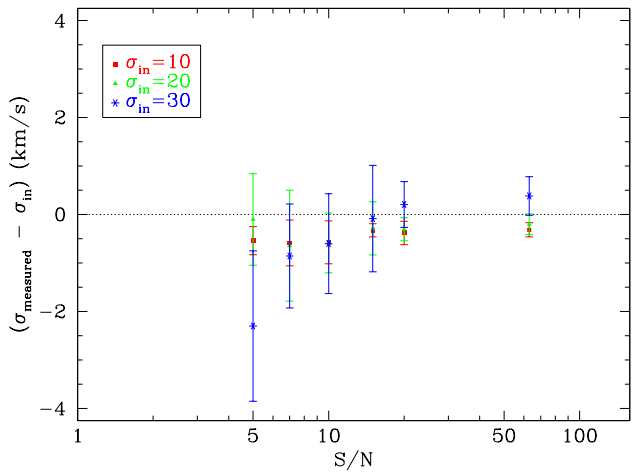}
}
\caption[]{Difference between the measured and simulated velocity dispersion 
as a  function of S/N of the spectra. These simulations are based on randomly 
selected 7 template stars whose spectra were broadened by 
convolving them with Gaussians with widths of 10, 20, and 30 km~s$^{-1}$ and 
had S/N degraded to simulate noisy
spectra, more similar to globular cluster targets. 
{\it (See the electronic edition of the
journal for the colour version of this figure.)}
}
\label{fig:diffsigSN}
\end{figure}

In Fig.~\ref{fig:diffsigSN} we tested the dependence of the measured velocity 
dispersion on S/N of the input spectra. The measured velocity dispersions
from the most broadened spectra, with
$\sigma_{in}=30$~km~s$^{-1}$ and with the lowest S/N, appear to be slightly
underestimated. However, these have, as expected higher uncertainty and are consistent
with the input values within the errors.

The observations of radial velocity standards were also 
secured during the 1993 EMMI run and were used to check that the CCF 
has Gaussian shape and that its width does not depend on the 
stellar metallicity \citep[see also][]{dubath+90,
dubath+92}. The average sigma of the stellar CCFs, derived from 10 
measurements of K-giant radial velocity standard stars observed during that 
same run, is $6.2\pm 0.3$~km~s$^{-1}$, where 
0.3~km~s$^{-1}$ is the standard deviation around the mean.


\subsection{Cluster radial velocities and velocity dispersions}
\label{sect:gcRV}

The initial estimate of the radial velocity for all the clusters was obtained by
fitting the $H_\alpha$ line. In all but one cluster spectrum 
the line was well defined and could be fitted with a Gaussian profile using 
{\it splot} task in IRAF. Then the precise radial velocities and velocity 
dispersions were measured using {\it fxcor} IRAF task.

\begin{table}
\caption[]{Radial velocity measurements for globular clusters observed with 
EMMI and UVES are listed in column (4). For clusters with multiple observations 
the reported value is a weighted average of velocities measured from individual 
spectra and the combined spectrum. The last column is the radial velocity from 
\citet{peng+04GCcat} catalogue. $B-V$ and $V-I$ colours were  
taken from \citet{peng+04GCcat} when available, otherwise from original 
discovery papers. They have been dereddened assuming only foreground reddening 
of E(B--V)=0.11 \citep{schlegel+98}, except for clusters observed 
by \citet[][HCH99]{holland+99} which have individual reddenings from that work. 
Reddening assumed for HGHH92-C23 is 0.31, derived from
strong interstellar NaD absorption lines (see text for more details).}
\label{tab:RVgc}
\begin{tabular}{lccll}
\hline \hline
\multicolumn{1}{c}{(1)} & \multicolumn{1}{c}{(2)} & \multicolumn{1}{c}{(3)}  &
\multicolumn{1}{c}{(4)}&\multicolumn{1}{c}{(5)}\\
\multicolumn{1}{c}{ID} & \multicolumn{1}{c}{$(B-V)_0$}&\multicolumn{1}{c}{$(V-I)_0$}& 
\multicolumn{1}{c}{V$_R$}  & \multicolumn{1}{c}{V$_R$(P04)} \\  
		&	 \multicolumn{1}{c}{(mag)}  & \multicolumn{1}{c}{(mag)}  &
\multicolumn{1}{c}{(km~s$^{-1}$)}  & \multicolumn{1}{c}{(km~s$^{-1}$)}\\
\hline
HGHH92-C1          &  \multicolumn{1}{c}{~$\cdots$~} & \multicolumn{1}{c}{~$\cdots$~} &$ 642.4 \pm 1.0 $& $ 633 \pm 11 $\\
VHH81-C3           & 0.91 & 1.08 &$ 561.8 \pm 1.6 $& $ 528 \pm 65 $\\
VHH81-C5           & 0.70 & 0.82 &$ 556.6 \pm 2.5 $& $ 556 \pm 19 $\\
HGHH92-C6          & 0.85 &  1.00 &$ 854.5 \pm 1.8 $& $ 828 \pm 65 $\\
HGHH92-C7          & 0.75 &  0.91 &$ 594.9 \pm 0.5 $& $ 617 \pm 10 $\\
HGHH92-C11         & 0.94 &  1.12 &$ 753.0 \pm 0.4 $& $ 755 \pm 11 $\\
HGHH92-C12=R281    & \multicolumn{1}{c}{~$\cdots$~} & \multicolumn{1}{c}{~$\cdots$~} &$ 440.4 \pm 0.3 $& $ 443 \pm  9 $\\
HHH86-C15=R226     & 0.89 &  1.03 &$ 644.3 \pm 0.4 $& $ 638 \pm 18 $\\
HGHH92-C17         & 0.77 &  0.88 &$ 781.3 \pm 1.8 $& $ 783 \pm 12 $\\
HHH86-C18          & 0.78 &  0.92 &$ 479.8 \pm 2.1 $& $ 494 \pm 65 $\\
HGHH92-C21         & 0.78 &  0.93 &$ 461.3 \pm 2.1 $& $ 465 \pm 7  $\\
HGHH92-C22         & 0.79 &  0.91 &$ 578.4 \pm 0.3 $& $ 565 \pm 13 $\\
HGHH92-C23         & 0.76 &  0.78 &$ 673.7 \pm 0.9 $& $ 677 \pm 9  $\\
HGHH92-C29         & 0.89 &  1.08 &$ 726.1 \pm 0.4 $& $ 733 \pm 10 $\\
HGHH92-C36=R113    & 0.73 &  0.85 &$ 702.7 \pm 1.1 $& $ 680 \pm 12 $\\
HGHH92-C37=R116    & 0.84 &  0.99 &$ 611.7 \pm 0.3 $& $ 630 \pm 14 $\\
HHH86-C38=R123     & 0.78 &  0.91 &$ 405.1 \pm 0.7 $& $ 418 \pm 12 $\\
HGHH92-C41         & 0.89 &  1.09 &$ 363.0 \pm 0.2 $& $ 370 \pm 15 $\\
HGHH92-C44         & 0.69 &  0.85 &$ 504.8 \pm 0.8 $& $ 538 \pm 56 $\\
HCH99-2            & 0.74 &  0.84 &$ 300.4 \pm 2.0 $& $ 299 \pm 18 $\\
HCH99-15           & \multicolumn{1}{c}{~$\cdots$~} &  1.06 &$ 518.6 \pm 0.7 $& \multicolumn{1}{c}{~$\cdots$~}\\
HCH99-16           & \multicolumn{1}{c}{~$\cdots$~} &  0.79 &$ 458.2 \pm 2.3 $& $ 454 \pm 44 $\\
HCH99-18           & 0.89 &  0.89 &$ 455.0 \pm 0.5 $& $ 447 \pm 14 $\\
HCH99-21           & \multicolumn{1}{c}{~$\cdots$~} &  0.78 &$ 662.9 \pm 1.5 $& $ 669 \pm 22 $\\
R122		   & \multicolumn{1}{c}{~$\cdots$~} & \multicolumn{1}{c}{~$\cdots$~} &$ 588.4 \pm 1.5 $&\multicolumn{1}{c}{~$\cdots$~}\\
R223		   & 0.80 &  0.95 &$ 775.7 \pm 0.6 $& $ 572 \pm 56 $\\
R261		   & 0.83 &  0.99 &$ 614.8 \pm 3.9 $& $ 613 \pm 12 $\\
\hline
\end{tabular}
\end{table}

Radial velocity measurements for all the clusters are listed in 
Table~\ref{tab:RVgc}. After the identifier, de-reddened $(B-V)_0$ and
$(V-I)_0$ colours of the targets are given. The radial velocities measured 
from our spectra are in column 4, and the velocities from the catalogue of 
\citet{peng+04GCcat} are shown for comparison in the last column.
For clusters with both EMMI and UVES spectra the radial velocities reported 
are weighted averages of both measurements. The individual radial velocity 
and velocity dispersion measurements for clusters with multiple observations 
are reported in Table~\ref{tab:multispec}. We note that the values listed as 
``combined'' are not the averages of individual measurements, but rather 
the measurements of the radial velocity and velocity dispersion from the
combined UVES spectra, constructed by averaging individual exposures. 
These spectra have slightly higher S/N and the agreement between the values 
obtained from the individual and these combined spectra indicates the absence of 
significant systematic errors (Fig.~\ref{fig:diffsigSN}).

\begin{table}
\caption[]{Radial velocities and velocity dispersions measured on individual
spectra for cluster targets with multiple observations. The values listed as 
``combined'' are not the averages of individual measurements, but rather 
the measurements from the
combined UVES spectra, constructed by averaging individual exposures.}
\label{tab:multispec}
\begin{tabular}{lccl}
\hline \hline
\multicolumn{1}{c}{(1)} & \multicolumn{1}{c}{(2)} & 
\multicolumn{1}{c}{(3)}  & \multicolumn{1}{c}{(4)}\\
\multicolumn{1}{c}{ID} & \multicolumn{1}{c}{V$_R$}  & 
\multicolumn{1}{c}{$\sigma$} & \multicolumn{1}{c}{Inst.} \\  
 & \multicolumn{1}{c}{km~s$^{-1}$}  & 
\multicolumn{1}{c}{km~s$^{-1}$} &  \\  
\hline
\multicolumn{4}{l}{HGHH92-c1}\\
combined & $  638.0 \pm 5.6$&$ 13.1 \pm 2.3$& UVES \\
A& $ 639.2 \pm 3.8 $&$ 11.9 \pm 1.2 $&UVES \\
B& $ 643.6 \pm 1.8 $&$ 14.1 \pm 1.4 $&UVES \\
1& $ 642.5 \pm 2.1 $&$ 14.1 \pm 1.7 $&EMMI \\
\hline
\multicolumn{4}{l}{HGHH92-c6}\\
1& $ 857.3 \pm 2.1 $&$ 19.5 \pm 1.7 $&EMMI  \\
2& $ 847.9 \pm 3.2 $&$ 24.5 \pm 3.0 $&EMMI \\
\hline
\multicolumn{4}{l}{HGHH92-c7}\\
combined& $ 592.9 \pm 1.4 $&$ 24.1 \pm 1.6 $&UVES \\
A& $ 594.7 \pm 0.9 $&$ 21.1 \pm 0.1 $&UVES \\
B& $ 595.8 \pm 0.7 $&$ 24.2 \pm 1.4 $&UVES \\
C& $ 595.1 \pm 0.9 $&$ 24.7 \pm 1.4 $&UVES \\
1& $ 590.9 \pm 2.3 $&$ 22.2 \pm 1.9 $&EMMI \\
2& $ 593.6 \pm 2.1 $&$ 17.6 \pm 1.8 $&EMMI \\
\hline
\multicolumn{4}{l}{HGHH92-c11}\\
combined& $ 753.1 \pm 0.5 $&$ 18.4 \pm 1.0 $&UVES \\
A& $ 752.4 \pm 1.2 $&$ 18.2 \pm 2.0 $&UVES \\
B& $ 753.1 \pm 0.8 $&$ 19.4 \pm 0.4 $&UVES \\
\hline
\multicolumn{4}{l}{HGHH92-c12 = R281}\\
A& $ 440.4 \pm 0.3 $&$ 13.1 \pm 0.5 $&UVES \\
1& $ 439.2 \pm 1.5 $&$ 13.4 \pm 0.9 $&EMMI \\
\hline
\multicolumn{4}{l}{HGHH92-c17}\\
1& $ 777.8 \pm 2.8 $&$ 16.5 \pm 2.5 $& EMMI \\
2& $ 783.7 \pm 2.3 $&$ 23.7 \pm 2.0 $& EMMI \\
\hline
\multicolumn{4}{l}{HGHH92-c21}\\
A& $ 460.2 \pm 1.2 $&$ 19.0 \pm 0.1 $&UVES \\
1& $ 465.5 \pm 2.4 $&$ 16.3 \pm 2.1 $&EMMI \\
\hline
\multicolumn{4}{l}{HGHH92-c23}\\
combined& $ 671.4 \pm 1.3 $&$ 30.9 \pm 1.5 $&UVES \\
A& $ 674.5 \pm 1.7 $&$ 30.5 \pm 0.2 $&UVES \\
B& $ 675.8 \pm 1.5 $&$ 28.6 \pm 2.1 $& UVES \\
1& $ 673.4 \pm 2.1 $&$ 26.3 \pm 1.8 $& EMMI \\
2& $ 675.8 \pm 2.9 $&$ 25.8 \pm 2.7 $& EMMI \\
\hline
\multicolumn{4}{l}{HGHH92-c44}\\
combined& $ 504.2 \pm 1.1 $&$  8.4 \pm 3.4 $& UVES \\
A& $ 505   \pm 18  $&$  9.8 \pm 2.2 $& UVES \\
B& $ 505.4 \pm 1.1 $&$ 14.6 \pm 1.2 $& UVES \\
\hline
\end{tabular}
\end{table}

\citet{dubath+97} have made detailed numerical simulations in order to 
understand and estimate the statistical errors on their radial velocity and 
projected velocity dispersion $\sigma_p$ measurements obtained by applying 
a cross-correlation technique to integrated-light spectra. 
They show that statistical errors, which can be very important for 
integrated-light measurements of Galactic (nearby)
globular clusters, because of the dominance of a few bright stars, are 
negligible in the present case where sampling problems are not present, thanks to the
larger distances of our targets. An integrated light spectrum of an NGC 5128 
globular cluster is well approximated by the convolution of the spectrum of 
a typical globular cluster star with the projected velocity distribution. 
The influence of binary stars is negligible 
\citep[e.g.][for dSph galaxies]{olszewski+96}.

The measurements of velocity dispersions for all the clusters are given in
Table~\ref{tab:Veldispgc}. In the first column is the identifier, then we list 
velocity dispersion measured on UVES spectra. In the third column the velocity
dispersions measured on EMMI spectra are given, while in the last column the
results from \citet{martini+ho04} are shown for comparison. 

\begin{figure*}
\centering
\resizebox{\hsize}{!}{
\includegraphics[angle=270]{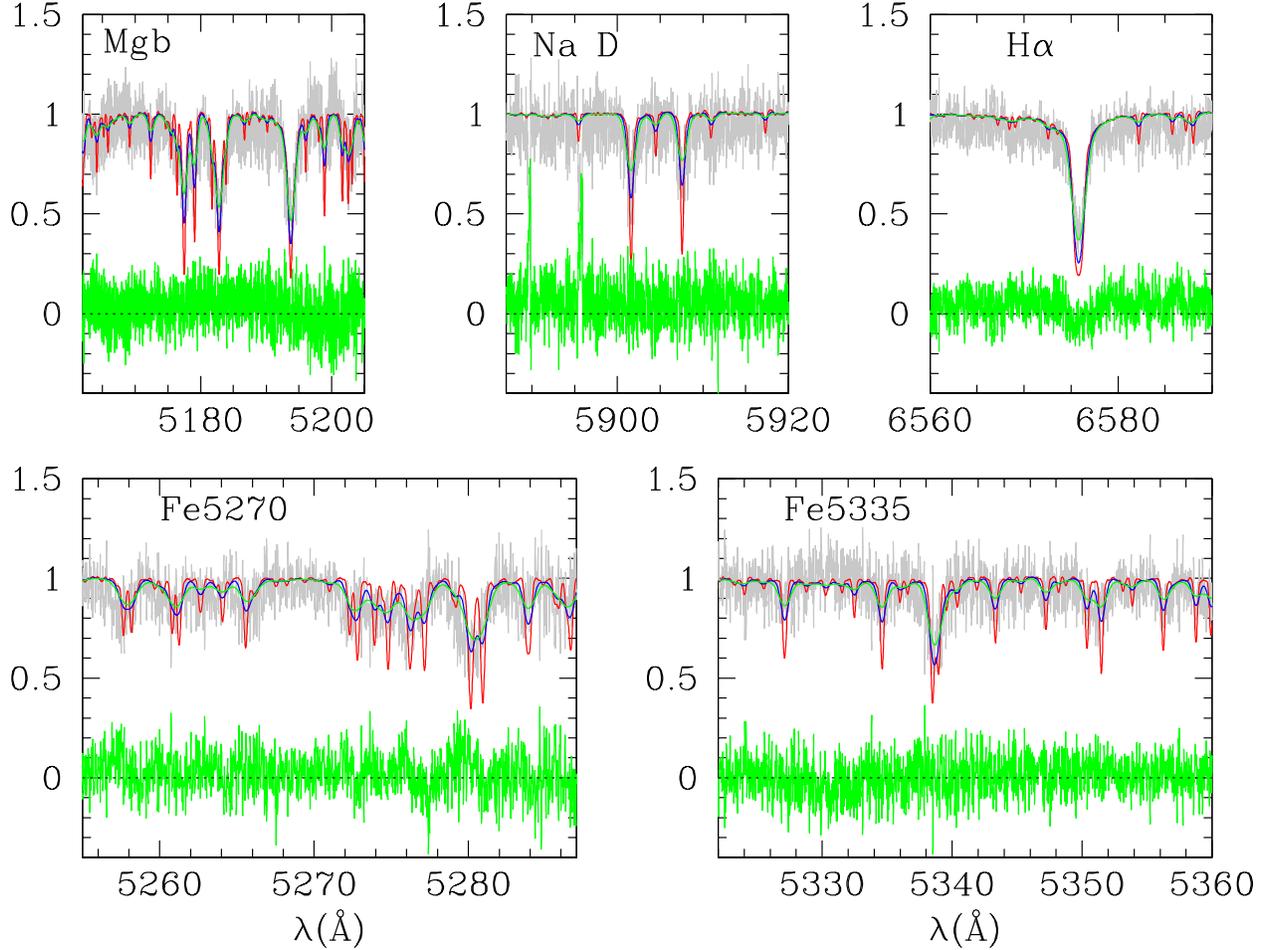}
}
\caption[]{The spectrum of HGHH92-C7 is plotted in gray showing the spectral
regions around some of the prominent absorption line features: Mgb, Na~D and
H$\alpha$ in the upper panel and Fe5270 and Fe5335 features in the lower panel.
Overplotted are spectra of HD~103295 star broadened to simulate velocity 
dispersion of 5 (red line), 15 (blue) and 25 km~s$^{-1}$ (green). The best 
fitting template is the one widened by 25~km~s$^{-1}$ in agreement with 
$24.1 \pm 1.6$~km~s$^{-1}$ obtained by cross-correlation 
(Table~\ref{tab:multispec}). The differences (template-GC)
are shown at the bottom of each panel demonstrating the goodness of the fit  
for the template broadened with $\sigma=25$~km~s$^{-1}$.
{\it (See the electronic edition of the
journal for the colour version of this figure.)}
}
\label{fig:C7spect}
\end{figure*}

In Fig.~\ref{fig:C7spect} we plot the spectrum of the highest S/N cluster, the
combined three 20 min exposures of HGHH92-C7, centered on some of the
characteristic absorption lines. Overplotted are
broadened spectra of HD~103295 made by convolving with Gaussians of 5 (red), 15
(blue) and 25~km~s$^{-1}$ (green). The best fitting template is the one broadened 
to 25~km~s$^{-1}$ in agreement with $24.1 \pm 1.6$~km~s$^{-1}$ obtained 
by cross-correlation (Table~\ref{tab:multispec}). The differences between the
template broadened with 25~km~s$^{-1}$ and the cluster spectra are shown below the
spectra in each panel. The narrow 
lines at $5890.0$ and $5895.9$~\AA, cannot be fitted by the
broadened stellar templates. This is expected, because they are resonance lines
(Na~D1 and Na~D2) due to interstellar ions of NaI.
The equivalent width of these lines can be
used to constrain the interstellar extinction towards each globular cluster
\citep{munari+zwitter97}. The equivalent width of Na~D1 line in this spectrum
implies $\mathrm{E(B}-\mathrm{V)} \simeq 0.07$~mag, somewhat lower than 
the average reddening of 0.11~mag
in the line of sight of NGC~5128 from \citet{schlegel+98} maps. However, we note
that both our measurement and the calibration have considerable uncertainty and
that the error on the reddening is probably of the same size as the derived value.

\begin{table}
\caption[]{Velocity dispersion measurements for the clusters 
observed with UVES are listed in column 2 and for those observed with 
EMMI in column 3. For clusters with several observations  
the reported value is the weighted average of individual measurements
(for each instrument separately). 
All the measurements from individual spectra are reported in
Table~\ref{tab:multispec}. 
For comparison in column 4 we list velocity dispersion measurements 
from \citet{martini+ho04}.}
\label{tab:Veldispgc}
\begin{tabular}{lrrr}
\hline \hline
\multicolumn{1}{c}{(1)} & \multicolumn{1}{c}{(2)} & \multicolumn{1}{c}{(3)}  & \multicolumn{1}{c}{(4)}\\
\multicolumn{1}{c}{ID} &  \multicolumn{1}{c}{$\sigma$(UVES)}      & \multicolumn{1}{c}{$\sigma$(EMMI)}
 & \multicolumn{1}{c}{$\sigma$(MH04)}  \\
                &   \multicolumn{1}{c}{(km~s$^{-1}$)}  &
\multicolumn{1}{c}{(km~s$^{-1}$)}&\multicolumn{1}{c}{(km~s$^{-1}$)} \\
\hline
HGHH92-C1      & $ 12.9 \pm 0.8 $              & $ 14.1 \pm 1.7 $              & \multicolumn{1}{c}{~$\cdots$~}\\
VHH81-C3       & \multicolumn{1}{c}{~$\cdots$~}& $ 16.1 \pm 1.1 $  	       & \multicolumn{1}{c}{~$\cdots$~}\\
VHH81-C5       & \multicolumn{1}{c}{~$\cdots$~}& $ 15.8 \pm 2.2 $  	       & \multicolumn{1}{c}{~$\cdots$~}\\
HGHH92-C6      & \multicolumn{1}{c}{~$\cdots$~}& $ 20.7 \pm 1.5 $  	       & \multicolumn{1}{c}{~$\cdots$~}\\
HGHH92-C7      & $ 21.1 \pm 0.1 $   	       & $ 19.8 \pm 1.3 $  	       & $ 22.4 \pm 2.1 $              \\
HGHH92-C11     & $ 19.2 \pm 0.4 $   	       & \multicolumn{1}{c}{~$\cdots$~}& $ 17.7 \pm 1.9 $              \\
HGHH92-C12=R281& $ 13.1 \pm 0.5 $	       & $ 13.4 \pm 0.9 $	       & \multicolumn{1}{c}{~$\cdots$~}\\
HHH86-C15=R226 & $ 11.1 \pm 0.7 $	       & \multicolumn{1}{c}{~$\cdots$~}& \multicolumn{1}{c}{~$\cdots$~}\\
HGHH92-C17     & \multicolumn{1}{c}{~$\cdots$~}& $ 20.9 \pm 1.6 $ 	       & $ 18.9 \pm 2.0 $              \\
HGHH92-C18     & \multicolumn{1}{c}{~$\cdots$~}& $ 15.3 \pm 1.8 $              & \multicolumn{1}{c}{~$\cdots$~}\\
HGHH92-C21     & $ 19.0 \pm 0.1 $	       & $ 16.3 \pm 2.1 $	       & $ 20.8 \pm 1.9 $	       \\
HGHH92-C22     & $ 17.9 \pm 0.1 $	       & \multicolumn{1}{c}{~$\cdots$~}& $ 19.1 \pm 2.0 $	       \\
HGHH92-C23     & $ 30.5 \pm 0.2 $	       & $ 26.1 \pm 1.5 $	       & $ 31.4 \pm 2.6 $	       \\
HGHH92-C29     & $ 16.1 \pm 0.8 $	       & \multicolumn{1}{c}{~$\cdots$~}& $ 16.1 \pm 2.1 $	       \\
HGHH92-C36=R113& $ 15.7 \pm 1.9 $	       & \multicolumn{1}{c}{~$\cdots$~}& \multicolumn{1}{c}{~$\cdots$~}\\
HGHH92-C37=R116& $ 12.6 \pm 0.8 $	       & \multicolumn{1}{c}{~$\cdots$~}& $ 13.5 \pm 1.6 $              \\
HHH86-C38=R123 & $ 14.2 \pm 1.1 $	       & \multicolumn{1}{c}{~$\cdots$~}& \multicolumn{1}{c}{~$\cdots$~}\\
HGHH92-C41     & $ 11.5 \pm 1.3 $	       & \multicolumn{1}{c}{~$\cdots$~}& $  9.6 \pm 2.0 $	       \\
HGHH92-C44     & $ 13.1 \pm 1.0 $	       & \multicolumn{1}{c}{~$\cdots$~}& $  9.1 \pm 2.0 $	       \\
HCH99-2        & $ 14.1 \pm 0.5 $	       & \multicolumn{1}{c}{~$\cdots$~}& \multicolumn{1}{c}{~$\cdots$~}\\
HCH99-15       & $ 21.3 \pm 1.7 $	       & \multicolumn{1}{c}{~$\cdots$~}& \multicolumn{1}{c}{~$\cdots$~}\\
HCH99-16       & $  9.5 \pm 1.4 $	       & \multicolumn{1}{c}{~$\cdots$~}& \multicolumn{1}{c}{~$\cdots$~}\\
HCH99-18       & $ 21.2 \pm 1.1 $	       & \multicolumn{1}{c}{~$\cdots$~}& \multicolumn{1}{c}{~$\cdots$~}\\
HCH99-21       & $ 10.6 \pm 2.3 $	       & \multicolumn{1}{c}{~$\cdots$~}& \multicolumn{1}{c}{~$\cdots$~}\\
R122	       & $  4.9 \pm 1.1: $	       & \multicolumn{1}{c}{~$\cdots$~}& \multicolumn{1}{c}{~$\cdots$~}\\
R223	       & $ 14.4 \pm 1.5 $	       & \multicolumn{1}{c}{~$\cdots$~}& \multicolumn{1}{c}{~$\cdots$~}\\
R261	       & $ 14.6 \pm 0.7 $	       & \multicolumn{1}{c}{~$\cdots$~}& \multicolumn{1}{c}{~$\cdots$~}\\
\hline
\end{tabular}
\end{table}

\begin{figure*}
\centering
\resizebox{\hsize}{!}{
\includegraphics[angle=270]{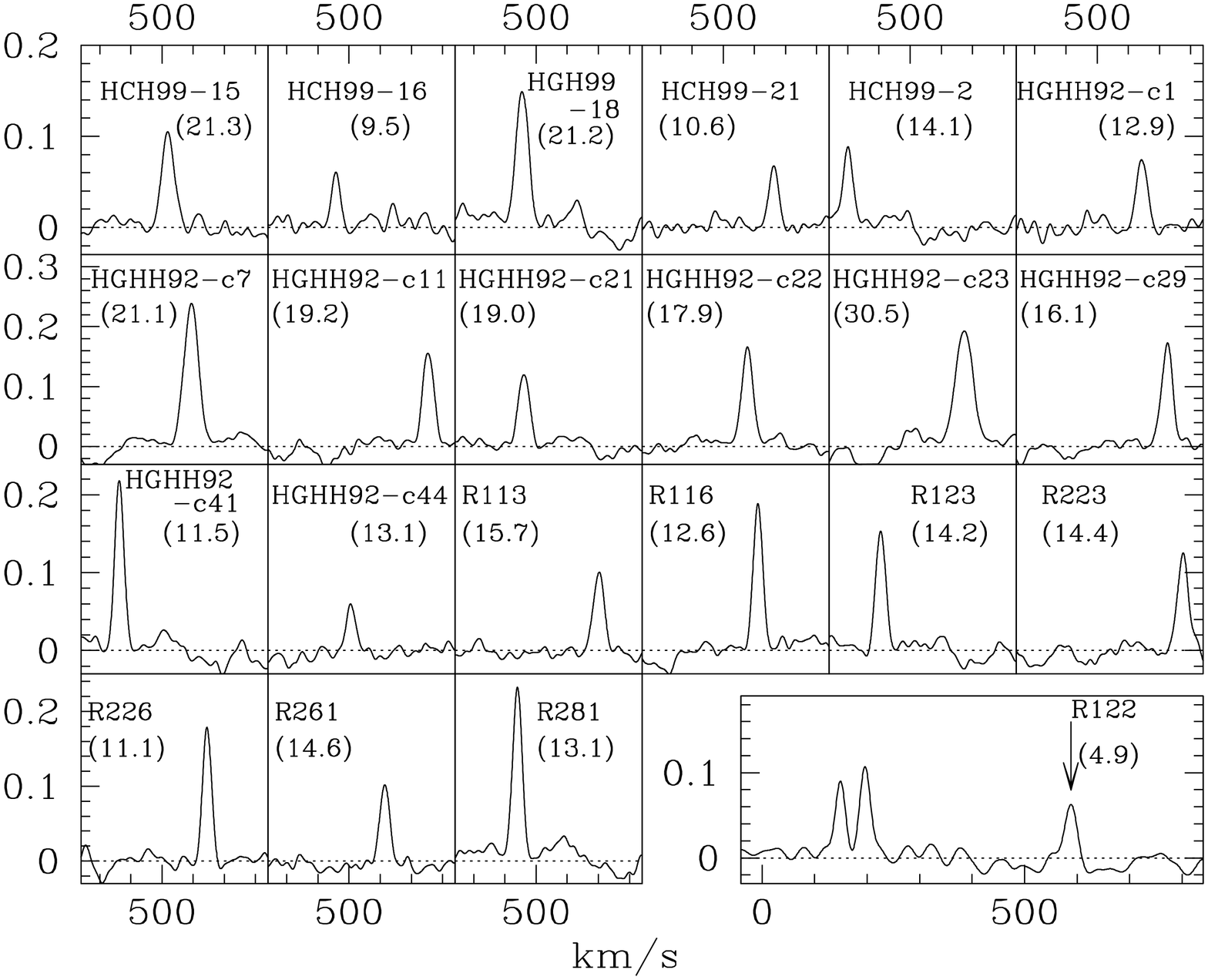}
}
\caption[]{The CCFs for all the clusters observed with UVES. The  
derived velocity dispersion is given in the parenthesis
next to the name of each cluster. The
x-axis scale has been corrected for the relative velocity shift between the
template star and the cluster so that the peak of the CCFs correspond to the
heliocentric velocity of the cluster. The x-axis displays one tick for every 
100 km~s$^{-1}$. All the CCFs show a single well defined
Gaussian peak, except for that of R122, which has additional peaks at velocities
59 and at 114 km~s$^{-1}$ due to contamination by Galactic foreground stars.
}
\label{fig:gcCCFs}
\end{figure*}

In Fig.~\ref{fig:gcCCFs} we plot cross-correlation functions for all the clusters.
In each panel, next to the cluster name, the velocity dispersion is given in 
parenthesis.  The
x-axis scale has been corrected for the relative velocity shift between the
template star and the cluster so that the peak of the CCFs correspond to the
heliocentric velocity of each cluster. 
All the CCFs show a single well defined
Gaussian peak, except for R122, which has additional peaks at velocities
59 and at 114 km~s$^{-1}$. We have inspected the through-slit image taken at the
start of the exposure as well as deep $V$-band images taken with FORS~1 under
excellent seeing conditions, but there is no sign of a spatially resolved
blend at this position. The
two additional peaks at the above given velocities are present when
cross-correlating the cluster spectrum with all the template stars and the height
of these peaks is comparable, and sometimes larger than that corresponding to a
velocity plausible for a cluster in NGC~5128. 
We note that the spectrum of this object also displays additional absorption
features. We do not have any explanation for this, other
than an unfortunate spatially unresolved 
blend with foreground MW stars, which is not uncommon
given the low Galactic latitude ($b=+19.4^\circ$) of  NGC5128.

To calculate the mean and sigma of the velocity dispersion for each spectrum, we
first calculate the straight mean and standard deviation of the 18 independent
measurements, each obtained by cross-correlating the cluster spectrum
with a different template star. The error is the quadratic sum of the standard
deviation and the uncertainty in the calibration of the instrumental width for the
CCF. This is done separately for the upper and lower
UVES CCD. We combine the two averaged velocity dispersion measurements
through the weighted mean. Since the two 
chips have slightly different resolution and the 
calibration of the $\sigma_{ref}$ is independently made, the uncertainty in the
mean is calculated with the average variance of the data using the following
expression \citep{bevington}:
\begin{equation}
\sigma_\mu = \sqrt{\frac{\sum\limits_i [w_i  (x_i - \mu)^2]}{(N-1) \sum\limits_i w_i}}
\end{equation}
where $w_i = 1/\sigma_i^2$ is the usual definition of weights and $\mu$ is the
weighted mean of the $x_i$ measurements from the two chips.

For clusters with multiple observations, the measurements of velocity 
dispersion from individual spectra are given in Table~\ref{tab:multispec}, 
while in Table~\ref{tab:Veldispgc} we list the weighted average of the 
velocity dispersion from all the measurements on spectra taken with the given
instrument. In particular the uncertainty in the mean is calculated according to
\citep{bevington}:
\begin{equation}
\sigma_\mu = \sqrt{\frac{1}{\sum\limits_{i}w_i}}
\end{equation}


\subsubsection{Comparison with previous measurements}
\label{sect:compare}

\begin{figure}
\centering
\resizebox{\hsize}{!}{
\includegraphics[angle=270]{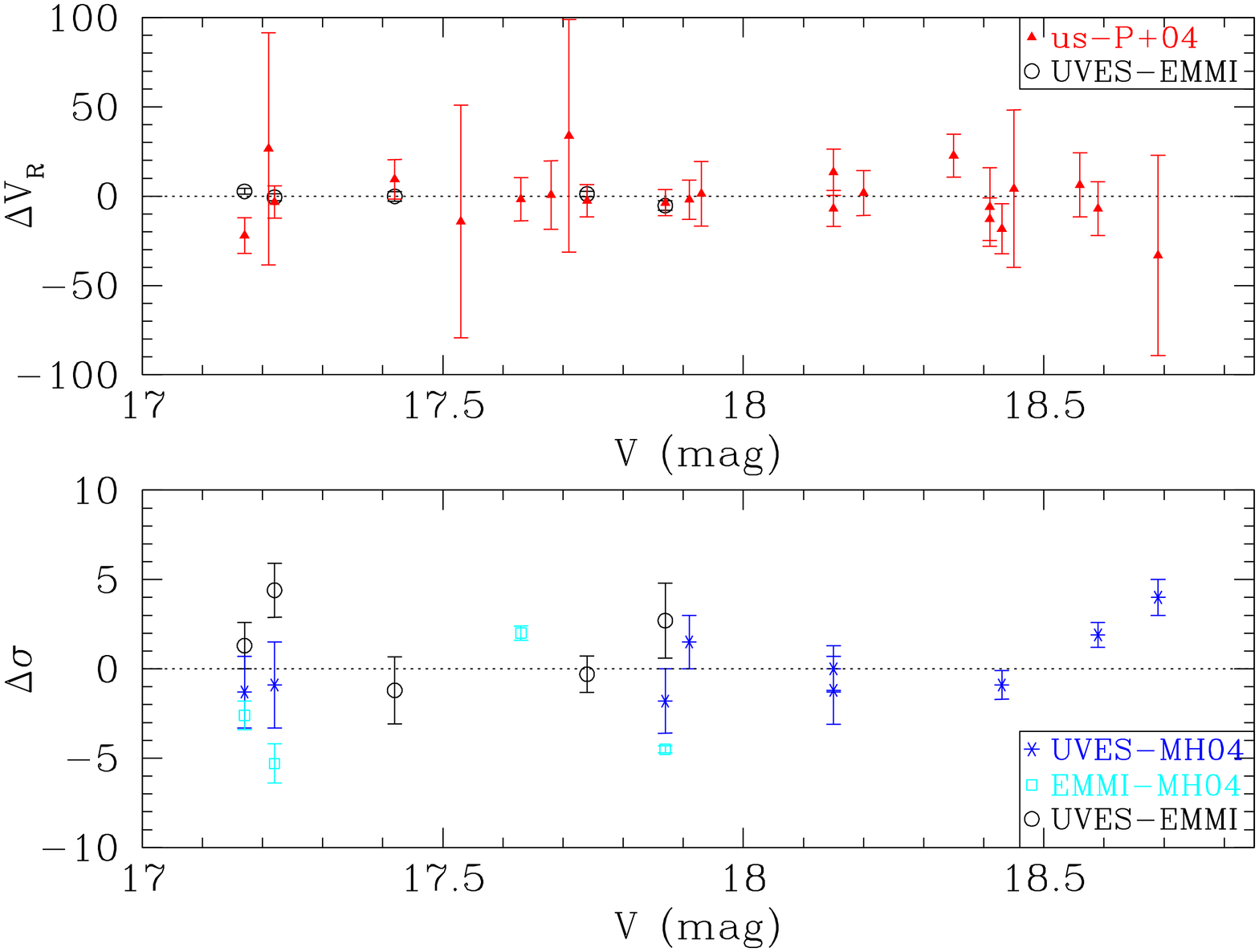}
}
\caption[]{Comparison of the measurements from the literature and our results, as
well as comparison between UVES and EMMI measurements for the clusters in common. 
In the upper panel we plot the difference between radial velocity and in the lower
panel velocity dispersion differences as a function of magnitude. In both panels
the open circles are used to compare UVES and EMMI measurements. The filled
triangles are comparing radial velocities with those of 
\citet[][P+04]{peng+04GCcat}. In the lower panel our
velocity dispersions are compared with those of \citet[][MH04]{martini+ho04} for
clusters observed with UVES (asterisks) and EMMI (open squares) separately. 
{\it (See the electronic edition of the
journal for the colour version of this figure.)}}
\label{fig:comparelit}
\end{figure}

In the upper panel of 
Fig.~\ref{fig:comparelit} we compare our radial velocity measurements with those
of \citet[][filled triangles]{peng+04GCcat}. The
average difference is $8 \pm 43$~km~s$^{-1}$. 
Due to high resolution and the wide wavelength
coverage the errors in radial velocities of our spectra are significantly 
smaller in spite of their relatively low S/N. The agreement with the previously
published measurements is excellent. 

The comparison between the velocity dispersions for the clusters in
common with \citet{martini+ho04} work is shown in the lower panel of
Fig.~\ref{fig:comparelit}. The average difference between our velocity 
dispersions from UVES spectra and those of \citet[][asterisks]{martini+ho04} 
is $0.14 \pm 1.93$~km~s$^{-1}$, while there is a slightly larger 
difference, amounting to $2.6 \pm 3.3$~km~s$^{-1}$, between the 
results obtained by these authors and our  
velocity dispersions obtained from EMMI spectra (open squares).

For the five clusters in common between our UVES and EMMI datasets, the average 
difference is $-0.4 \pm 3.0$~km~s$^{-1}$ 
for radial velocities and $1.4 \pm 2.3$~km~s$^{-1}$ for 
velocity dispersions. 
There is no trend with cluster brightness neither for radial velocity, nor for
velocity dispersion residuals.


\section{Globular cluster structural parameters}
\label{sect:structure}

\begin{table*}
\caption[]{Globular cluster structural parameters, projected galactocentric distances,  absolute magnitudes,
masses, and mass-to-light measurements. The columns are: (1) ID, (2) core radius ($r_c$), 
(3) projected two-dimensional effective (half-light) radius ($r_e$), 
(4) three-dimensional half-mass radius ($r_h$), 
(5) concentration ($c=\log r_t/r_c$), (6) ellipticity ($\epsilon$), (7) projected galactocentric 
distance ($R_{gc}$), (8)  absolute $V$ magnitude,
(9) virial mass ($M_{vir}$) in $10^6$~M$\sun$, and (10) $M/L_V$. 
In the last column we list the source for the structural parameters.}
\label{tab:structpar}
\begin{tabular}{lrrrrlrrrrl}
\hline \hline
\multicolumn{1}{c}{(1)}& \multicolumn{1}{c}{(2)}  & \multicolumn{1}{c}{(3)}   & \multicolumn{1}{c}{(4)} & 
\multicolumn{1}{c}{(5)}             & \multicolumn{1}{c}{(6)}        & \multicolumn{1}{c}{(7)} &
\multicolumn{1}{c}{(8)}& \multicolumn{1}{c}{(9)}& \multicolumn{1}{c}{(10)}& \multicolumn{1}{c}{(11)}\\
\multicolumn{1}{c}{ID} & \multicolumn{1}{c}{$r_c$}& \multicolumn{1}{c}{$r_e$} & \multicolumn{1}{c}{$r_h$} &
\multicolumn{1}{c}{$c$} & \multicolumn{1}{c}{$\epsilon$} & \multicolumn{1}{c}{$R_{gc}$} &\multicolumn{1}{c}{$M_V$}
&\multicolumn{1}{c}{$M_{vir}$}& 
\multicolumn{1}{c}{$M/L_V$} & \multicolumn{1}{c}{Ref.}\\
                       & \multicolumn{1}{c}{(pc)} & \multicolumn{1}{c}{(pc)}  &\multicolumn{1}{c}{(pc)} &
\multicolumn{1}{c}{($\log r_t/r_c$)}&                    & \multicolumn{1}{c}{(kpc)}& \multicolumn{1}{c}{(mag)}
& \multicolumn{1}{c}{($10^6$~M$\sun$)}& 
\multicolumn{1}{c}{(M$\sun$/L$\sun$)} &               \\
\hline
HGHH92-C7      &$1.4 \pm 0.1$ &$ 7.5 \pm 0.1$& $10.0 \pm 0.2$ & 1.83 & 0.13  & 9.3  &   $-11.09$& $ 7.8^{+0.7}_{-1.9}$&$ 3.3^{+0.8}_{-1.1}$ & 2 \\
HGHH92-C11     &$1.3 \pm 0.1$ &$ 7.8 \pm 0.1$& $10.4 \pm 0.2$ & 1.88 & 0.26  & 6.7  &   $-10.35$& $ 6.7^{+0.6}_{-1.6}$&$ 5.7^{+1.4}_{-1.9}$ & 2 \\
HGHH92-C12=R281&$1.2 \pm 0.2$ &$10.8 \pm 1.4$& $14.4 \pm 1.9$ & 2.4  & 0.20  & 11.4 &   $-10.52$& $ 4.3^{+0.7}_{-1.3}$&$ 3.1^{+0.9}_{-1.2}$ & 1 \\
 HHH86-C15=R226&$1.2 \pm 0.2$ &$ 5.3 \pm 0.7$& $ 7.1 \pm 0.9$ & 1.9  & 0.05  & 11.9 &   $ -9.70$& $ 1.5^{+0.2}_{-0.5}$&$ 2.3^{+0.6}_{-0.9}$ & 1 \\
HGHH92-C17     &$2.3 \pm 0.1$ &$ 5.7 \pm 0.1$& $ 7.6 \pm 0.2$ & 1.43 & 0.07  & 6.2  &   $-10.63$& $ 5.8^{+0.5}_{-1.4}$&$ 3.8^{+0.9}_{-1.3}$ & 2 \\
HGHH92-C21     &$1.2 \pm 0.1$ &$ 7.0 \pm 0.1$& $ 9.3 \pm 0.2$ & 1.86 & 0.33  & 7.3  &   $-10.39$& $ 5.8^{+0.5}_{-1.4}$&$ 4.8^{+1.1}_{-1.6}$ & 2 \\
HGHH92-C22     &$1.1 \pm 0.1$ &$ 3.8 \pm 0.1$& $ 5.1 \pm 0.2$ & 1.62 & 0.09  & 5.8  &   $-10.11$& $ 2.8^{+0.3}_{-0.7}$&$ 3.0^{+0.7}_{-1.0}$ & 2 \\
HGHH92-C23     &$0.9 \pm 0.1$ &$ 3.3 \pm 0.1$& $ 4.4 \pm 0.2$ & 1.67 & 0.14  & 5.8  &   $-11.66$& $ 7.2^{+0.7}_{-1.8}$&$ 1.8^{+0.5}_{-0.6}$ & 2 \\
HGHH92-C29     &$1.2 \pm 0.1$ &$ 6.9 \pm 0.1$& $ 9.2 \pm 0.2$ & 1.87 & 0.11  & 21.2 &   $-10.11$& $ 4.1^{+0.4}_{-1.0}$&$ 4.4^{+1.0}_{-1.4}$ & 2 \\
HGHH92-C36=R113&$0.7 \pm 0.3$ &$ 3.6 \pm 0.3$& $ 4.8 \pm 0.4$ & 2.0  & 0.32: & 13.1 &   $ -9.91$& $ 2.0^{+0.3}_{-0.6}$&$ 2.6^{+0.6}_{-0.9}$ & 1 \\
HGHH92-C37=R116&$0.6 \pm 0.1$ &$ 3.3 \pm 0.1$& $ 4.4 \pm 0.2$ & 1.87 & 0.02  & 12.1 &   $ -9.83$& $ 1.2^{+0.1}_{-0.3}$&$ 1.7^{+0.4}_{-0.6}$ & 2 \\
HGHH92-C37=R116&$0.7 \pm 0.2$ &$ 2.9 \pm 0.3$& $ 3.9 \pm 0.4$ & 1.9  & 0.21: & 12.1 &   $ -9.83$& $ 1.1^{+0.1}_{-0.3}$&$ 1.5^{+0.4}_{-0.6}$ & 1 \\
 HHH86-C38=R123&$0.5 \pm 0.2$ &$ 2.8 \pm 0.2$& $ 3.7 \pm 0.3$ & 2.2  & 0.10  & 14.0 &   $ -9.85$& $ 1.3^{+0.2}_{-0.4}$&$ 1.8^{+0.4}_{-0.6}$ & 1 \\
HGHH92-C41     &$0.8 \pm 0.1$ &$ 4.5 \pm 0.1$& $ 6.0 \pm 0.2$ & 1.87 & 0.05  & 23.4 &   $ -9.67$& $ 1.4^{+0.1}_{-0.3}$&$ 2.2^{+0.5}_{-0.7}$ & 2 \\
HGHH92-C44     &$1.3 \pm 0.1$ &$ 5.7 \pm 0.1$& $ 7.6 \pm 0.2$ & 1.70 & 0.06  & 20.4 &   $ -9.57$& $ 2.3^{+0.2}_{-0.6}$&$ 3.9^{+0.9}_{-1.3}$ & 2 \\
HCH99-2        &$1.0 \pm 0.1$ &$11.4 \pm 1.1$& $15.2 \pm 1.5$ & 1.5  & 0.08  & 2.6  &   $-10.33$& $ 5.3^{+0.7}_{-1.5}$&$ 4.5^{+1.2}_{-1.6}$ & 3 \\
HCH99-15       &$1.6 \pm 0.1$ &$ 5.9 \pm 0.2$& $ 7.8 \pm 0.3$ & 1.0  & 0.08  & 1.2  &   $-10.82$& $ 6.2^{+0.6}_{-1.5}$&$ 3.4^{+0.8}_{-1.1}$ & 3 \\
HCH99-16       &$0.8 \pm 0.2$ &$12.1 \pm 0.6$& $16.1 \pm 0.8$ & 1.6  & 0.30  & 1.8  &   $-10.08$& $ 2.5^{+0.3}_{-0.6}$&$ 2.8^{+0.7}_{-0.9}$ & 3 \\
HCH99-18       &$1.2 \pm 0.0$ &$13.7 \pm 0.3$& $18.3 \pm 0.3$ & 1.5  & 0.03  & 1.5  &   $-11.38$& $14.3^{+1.3}_{-3.5}$&$ 4.7^{+1.2}_{-1.6}$ & 3 \\
HCH99-21       &$2.6 \pm 1.1$ &$ 7.1 \pm 2.7$& $ 9.5 \pm 3.6$ & 0.8  & 0.02  & 3.0  &   $-10.28$& $ 1.9^{+0.7}_{-1.0}$&$ 1.7^{+0.7}_{-1.0}$ & 3 \\
R122	       &$0.6 \pm 0.3$ &$ 2.2 \pm 0.4$& $ 2.9 \pm 0.4$ & 1.7  & 0.39: & 12.0 &   $-10.17$& $0.12^{+0.02}_{-0.04}:$&$0.12^{+0.03}_{-0.04}:$ & 1 \\
R223	       &$0.5 \pm 0.4$ &$ 2.6 \pm 0.3$& $ 3.5 \pm 0.4$ & 2.0  & 0.06  & 6.6  &   $ -9.49$& $ 1.3^{+0.2}_{-0.4}$&$ 2.3^{+0.6}_{-0.9}$ & 1 \\
R261	       &$0.4 \pm 0.2$ &$ 1.9 \pm 0.4$& $ 2.6 \pm 0.5$ & 2.0  & 0.17  & 8.2  &   $-10.06$& $ 1.0^{+0.2}_{-0.3}$&$ 1.1^{+0.3}_{-0.4}$ & 1 \\
\hline										      
\multicolumn{8}{l}{References: 1: FORS1 data (this work); 2: \citet{harris+02}; 3: \citet{holland+99}.}
										      
\end{tabular}									      
\end{table*}

\citet{holland+99} published the measurements of structural parameters for 21 globular
cluster candidates in the inner part of Cen~A, based on WFPC2 images from the 
Hubble Space Telescope (HST). \citet{harris+02} have used HST STIS unfiltered 
images to increase the number of clusters with
measured structural parameters to 43 in this galaxy. Most of the clusters in our sample
have the structural parameters available from these two works. However, for 5 clusters
observed with EMMI and 8 clusters observed with UVES such data do not exist in the
literature.

The globular cluster candidates selected for the spectroscopic observations from the 
list of \citet{rejkuba01} have, however, images taken with FORS1 imaging spectrograph 
at the ESO VLT under superb seeing conditions. We have thus used these 
images to derive the structural parameters for 8 clusters in our sample. Only one of 
them, R116 which is the same as HGHH92-C37, has previously measured parameters from 
HST imaging. The comparison between the derived parameters from 
the ground and the space data for this cluster (Table~\ref{tab:structpar}) 
shows good agreement (but in the further analysis we use the more accurate 
results from the HST data for this cluster), and lends confidence in the 
results from the King-profile fitting from these ground based images. 
Measurements of the profiles of the other 7 clusters observed with FORS1 are 
published here for the first time.

The full description of the FORS1 dataset used here is given by 
\citet{rejkuba01}  and we will thus not repeat it. The most relevant 
parameters for the profile measurements are the seeing and the pixel scale. 
The seeing measured on the deep combined U-band images was $0 \farcs 52$ and 
$0 \farcs 55$ for the field~1 and 2, respectively. The V-band images had 
seeing of $0 \farcs 54$ and $0 \farcs 46$, but unfortunately these bright 
globular clusters had saturated cores in V-band. The pixel scale is 
$0\farcs2/\mathrm{pix}$. 

To measure the structural parameters we have used ISHAPE programme 
\citep{larsen99,larsen01} which models the light distribution of the 
cluster by convolving the assumed analytical model of the intrinsic 
luminosity profile of the cluster with the stellar PSF. The convolved model 
image is then subtracted from the observed image of the cluster and, via a 
$\chi^2$ minimization algorithm, ISHAPE returns the best fitting model 
parameters and also produces the residual image which can be examined 
in detail.

In particular for the modeling of these globular clusters the King profile 
\citep{king62} was assumed. In this model the core and tidal radii of a cluster 
are defined by the concentration parameter $c = \log r_t/ r_c$. The implementation 
of the concentration parameter in ISHAPE is slightly different with its 
definition in 
the linear scale ($C = r_t/ r_c$). We shall call this ISHAPE concentration parameter 
$C$ in order to avoid confusion. Since $C$ is the most 
uncertain of the fitted parameters in ISHAPE \citep{larsen01} and its best fitting 
value depends on the initial guesses, we have run a series of measurements with the 
fixed $C$ of  5, 15, 30, 50, 75, 100, 150, 200, 250 and 300. In all cases the 
best fit (the lowest $\chi^2$) shows the smallest amount of residuals in the 
subtracted image. 

Table~\ref{tab:structpar} lists the structural parameters for all our targets, 
except for the 6 clusters for which no high enough resolution optical images were 
available. In column 1 we list the cluster ID and its core radius 
($r_c$) in column 2. Columns 3 and 4 report the projected two-dimensional 
half-light (effective) radius ($r_e$) and the three-dimensional half-mass radius 
($r_h$), respectively. Concentration parameter ($c$) is listed in column 5 and
ellipticity in column 6. In 
all cases cluster radii are given in parsecs assuming the distance of 3.84~Mpc 
\citep{rejkuba04}. This is the same distance used in the work of \citet{martini+ho04} 
allowing a straight-forward comparison. We note that using the shorter 
distance of \citet{ferrarese+06}, the radii would be 11\% smaller.

Out of 6 clusters which have their structural parameters determined
here for the first time, three have ellipticities larger than 0.1. We note
however that significantly larger ellipticity has been derived for cluster
HGHH92-C37 (R116) from our FORS1 data with respect to work of
\citet{harris+02}. The difference in this particular case might be due to the
location of the cluster close to the edge of FORS1 field, where image
distortions could have affected the measurement. The very high ellipticity of
R122 could be due to the fact that the image of this cluster is most probably
blended with some foreground source (see also above). We have put ":" sign
next to the ellipticity determinations from FORS1 data that are more uncertain.

\section{Cluster masses and mass-to-light ratios}
\label{sect:ML}

The masses and the mass-to-light ratios for all the clusters with the available 
(or new) structural parameter measurements are given in Table~\ref{tab:structpar}. 
The listed masses are virial masses calculated using the virial 
theorem in the form \citep{spitzer87}:
\begin{equation}
M_{vir} \simeq 2.5 \frac{3\sigma^2 r_h}{G}
\label{eq:virmass}
\end{equation}
where, assuming an isotropic velocity distribution, 
$3\sigma^2$ is the mean square velocity of the 
stars and the cluster half-mass radius ($r_h$) is related to the half-light
(effective) radius ($r_e$) through $r_e \approx 3 r_h / 4$ \citep{spitzer87}.

1" slit  at the distance of 3.84~Mpc \citep{rejkuba04} corresponds 
to 18.62 pc. Seeing better than $0\farcs8$ ensures that  most of 
the light is in the slit. 
 The central velocity dispersion $\sigma_0$ has been estimated from the King 
profile fits convolved with the Gaussian of the width that
reproduces the intensity profile with FWHM as measured along the spatial
direction in the slit for each target 
\citep[e.g.][]{djorgovski+97,hasegan+05}. The corrections from the observed to
central $\sigma$ range from 4--10\%, with the average correction of
$\sim 6$\% being valid for most of the clusters. 

Core radii of Cen~A clusters are typically smaller 
than the resolution element, even for space based imagers, and thus
the quoted  errors for $r_c$ in 
Table~\ref{tab:structpar} are probably underestimated. In addition, as stated
before, the concentration parameter is also relatively uncertain, especially for the
clusters that had structural parameters determined with ISHAPE. Therefore we 
prefer to use the virial mass estimator as described above, rather than 
deriving the masses using the King model approximation, which would imply using
central velocity dispersion, core radius, and concentration parameter
\citep{queloz+95}. The uncertainty in structural parameters imply also that the
corrections of the observed $\sigma$ to the central value of 
velocity dispersion and
to the infinite aperture $\sigma$ are quite uncertain. In the next section, when we
plot $\sigma_0$, we apply a flat average correction of 6\% for all the clusters. 

In the calculation of the virial masses (Equ.~\ref{eq:virmass}) 
$\sigma$ is the mean value of 
velocity dispersion averaged over the whole cluster. 
In all our targets 1" slit samples the light to at least 3.5 $r_c$, and in most
cases beyond 6~$r_c$. Comparing with $\omega$~Cen and 47~Tuc
\citep{meylan+95,mclaughlin+06}, the faint surface brightness and low velocity
dispersion have negligible contribution beyond 3-5 $r_c$, and consequently 
$\sim 90-95$\% of the light of the clusters is in the slit.  
The corrections from the observed $\sigma$ to infinite apertures 
estimated using the seeing convolved King profiles, are of the order of 
few percent. We prefer not to apply these rather uncertain 
corrections, but rather use the observed $\sigma$, since the corrections are 
smaller than the uncertainty on our other parameters. The negative error-bars
for mass and M/L ratios in Table~\ref{tab:structpar} include the maximum
estimated aperture correction.

As expected from the comparison of the velocity dispersion measurements, the 
derived masses for the clusters in common with the \citet{martini+ho04} 
sample are
in good agreement with their virial masses for the targets in common. With
more bright globular clusters our sample has a larger number of clusters 
with masses similar to, or larger than, the most
massive Milky Way cluster $\omega$~Cen 
\citep[$M_{virial}=3 \times 10^6$~M\sun][]{meylan+95} and G1 in M31
\citep[$M_{virial}=7.3 \times 10^6$~M\sun][]{meylan+01}. 

The mass-to-light ratios ($M/L_V$) are computed by dividing the derived masses 
with the $V$-band luminosities ($L_V$) which are calculated assuming the 
absolute $V$ magnitude of the Sun to be $M_{V,\odot}=+4.83$~mag:
\begin{equation}
L_V = 10^{[-0.4*(V - (m-M)_V - A_V -M_{V,\odot})]}
\end{equation}

The derived $M/L_V$ ratios for our sample of globular clusters range from 
0.1 to 5.9. However, we note that the cluster with the lowest $M/L_V$, R01-122, 
is the one that displays contribution from (perhaps) stellar contaminants in its high
resolution spectra (Fig.~\ref{fig:gcCCFs}). 
Thus its luminosity is expected to be overestimated, which would
then underestimate the $M/L_V$ ratio. Excluding this cluster, the 
smallest $M/L_V$ is 
1.1 and the average is $<M/L_V>=2.9 \pm 1.4$, like also observed 
by \citet{martini+ho04}.
These authors note that this value is larger than the average $M/L_V$ 
of globular clusters
in the Local Group galaxies and they explore the possible
explanations for this, concluding that the difference is most probably real.
Our analysis confirms their results. We discuss this further in
Sect.~\ref{sect:FP}.

The errors of in the mass and $M/L_V$ determinations reported in
Table~\ref{tab:structpar} include the errors from the velocity dispersion 
measurements and half-mass radii, and the errors due to aperture corrections, 
but do not include any systematic errors 
due to modeling or assumptions on reddening and distance.
We note that, assuming a smaller distance modulus to NGC~5128 as determined by
\citet[][]{ferrarese+06}, the $M/L_V$ ratios would actually increase  by 
12\% and increase the difference with respect to Local Group globular clusters. 

The presence of a significant internal reddening in NGC~5128 would have the
opposite effect. However, from the measurements of the equivalent widths of 
the interstellar NaD doublet we estimate that the extinction is consistent 
with very little or no internal reddening within the galaxy except 
for HGHH93-C23 whose spectra display multiple and stronger NaD absorption lines 
(at different velocities). Furthermore for the inner clusters, which are 
expected to suffer the most dust reddening, the internal reddening values 
derived by \citet{holland+99}, are in all cases smaller than 
$E(B-V)=0.14$~mag, and mostly below 0.1~mag. Unfortunately the uncertainty of 
the relation between NaD and equivalent widths and
$E(B-V)$ coupled with our noisy spectra, which
yield high uncertainty in the NaD line equivalent width measurements, 
does not allow us to determine accurate reddening directly from the spectra. 
For the inner clusters we adopt the 
de-reddened magnitudes from \citet{holland+99} 
for the computation of the $M/L_V$ ratios (Table~\ref{tab:structpar}). 
For HGHH93-C23 we assume an additional $E(B-V)=0.2$~mag due to 
dust internal to NGC~5128, while for all the other clusters only the foreground 
Milky Way extinction of $A_V=0.34$~mag is assumed \citep{schlegel+98}.

\section{Special clusters}

\subsection{Clusters with X-ray sources}
\label{sect:CX}

\citet{kraft+01}, \citet{minniti+04}, \citet{peng+04GCcat}, and
\citet{voss+gilfanov06} have studied the Chandra X-ray point sources
matching NGC5128 globular clusters. \citet{minniti+04}
concluded that X-ray sources tend to be located in redder clusters
(more metal-rich) and also preferentially reside in more luminous (massive)
globular clusters.  Three of the clusters in the present sample have
been flagged as X-ray point sources based on the Chandra observations.
These are C23 with luminosity $L_X=1.10\times10^{38}$, C21 with
$L_X=1.79\times10^{37}$, and C7 with $L_X=1.89\times10^{37}$. 
Clusters C7 and C23 are the second and the third most massive, and C21
is sixth most massive cluster in our sample. According
to the data of \citet{peng+04GCcat}, these three clusters are red, with
$V-I=1.1-1.3$, and luminous, with $V=17.9$ to $17.2$ (or $M_V=-10.3$ to $-11$).
The X-ray luminosities of C21 and C7 are expected from typical  
low-mass X-ray binaries (LMXBs).
The C23 luminosity puts this cluster on the bright tail of the distribution
of X-ray point sources in NGC~5128 globular clusters shown
in Figure 5 of \citet{minniti+04}. This can be due to the presence
of a couple of LMXBs in this massive cluster, or to a single ultra-compact
binary \citep{bildsten+deloye04}. The alternative explanation of a more
massive accreting BH cannot be excluded, but it is really not demanded by
the available data.  In this respect, C23 appears to be similar to
the globular cluster Bo375 of M31, which contains the brightest X-ray
point source in a spectroscopically confirmed globular cluster, with
$L_X=2-6\times10^{38}$ \citep{distefano+02}.

\subsection{The most massive cluster}
\label{sect:CM}

\begin{table}
\caption[]{ Properties of HCH99-18, $\omega$ Cen and G1. The data for the last
two clusters are from \citet{meylan+95,meylan+01} and \citet{harris96}. 
The two $(V-I)$ colours of HCH99-18 are from HST photometry of \citet{holland+99}, and in parenthesis 
the measurement from  \citet{peng+04GCS} catalogue. 
The reddening towards this cluster is separated into foreground Galactic reddening \citep{schlegel+98}, and
internal reddening within NGC~5128 along the line of sight \citep{holland+99}.}
\label{tab:HCH99-18}
\begin{tabular}{lrrr}
\hline \hline
&\multicolumn{1}{c}{HCH99-18} & \multicolumn{1}{c}{$\omega$~Cen} & \multicolumn{1}{c}{G1}  \\
\hline
$M_V$ (mag)          & $-11.38$       & $-10.29$ & $-10.94$  \\
$(B-V)$ (mag)        & $1.06$         & $0.78$   & $0.84$  \\
$(V-I)$ (mag)        & $0.99~~(1.50)$  & $1.05$   &   \\
$E(B-V)$ (mag)       & $0.11+0.06$    & $0.12$   & $0.06$  \\
$\sigma_p(0)$ (km/s) & $21.2$         & $22$     & $27.8$  \\
$r_c$ (pc)           & $1.2$         & $4.6$    & $0.52$  \\
$r_h$ (pc)           & $18.3$        & $13$     & $14$  \\
$c=\log r_t/ r_c$    & $1.5$          & $1.23$   & $2.5$  \\
$\epsilon$           & $0.030$        & $0.121$  & $0.2$  \\
$M_{virial}$ (M\sun) &$1.4\times 10^7$& $3.0 \times 10^6$&$7.3 \times 10^6$ \\
$M/L_V$              & $4.7$          & $2.4$    & $3.6$  \\
$R_{gc}$ (kpc)       & $1.5$          & $6.4$    & $40.1$ \\
$\mathrm{[Fe/H]}$    &                & $-1.62$  & $-0.95$ \\
\hline
\end{tabular}
\end{table}

The brightest cluster of our sample is HCH99-18, with $V=17.07$. This
is also the most massive cluster, with $M_{vir}=1.4\times 10^7$~M\sun,
and the largest cluster in size (Table 6). It is apparently a metal-rich cluster, 
which has normal infrared colours \citep{minniti+96}.  
It is located in the inner parts of NGC~5128, {only 1.5 kpc away from the
galactic nucleus,} where reddening might be a problem.
The internal reddening due to dust in NGC~5128 in the line of sight to this
cluster is $E(V-I)=0.1\, \mathrm{(i.e.\ } E(B-V)=0.06)$~mag \citep{holland+99}. 
This relatively low internal reddening is confirmed by the low total 
equivalent width of interstellar NaD lines.
Its $M/L_V$ ratio is 4.7, higher than those of
$\omega$~Cen ($M/L_V=2.4$) and G1 ($M/L_V=3.6$)
globular clusters where the masses and $M/L$ were always taken from 
the same method, virial theorem \citep{meylan+01}.  The  $(V-I)$ 
colour of 
HCH99-18 taken from the original discovery work of \citet{holland+99} is 0.99,
significantly bluer than the $(V-I)=1.50$ in \citet{peng+04GCS} catalogue.

For a straight forward comparison of this 
most massive cluster (so far) in NGC~5128, with the $\omega$ Cen and G1 we
summarize their main properties in Table~\ref{tab:HCH99-18}. Data for 
$\omega$~Cen and G1 are from \citet{meylan+95,meylan+01}
and \citet{harris96}. There is no available literature value for the
spectroscopy of HCH99-18, but its redder colours indicate slightly 
higher metallicity if the same old age is assumed as for the other two 
clusters. This could explain the higher $M/L_V$ value of 
this cluster. Its mass is factor of 2 larger than that of G1, the more
massive of the two Local Group clusters. 

As already mentioned in the introduction, one of the favoured formation 
scenarios for $\omega$ Cen and G1 is being remnant nucleus of stripped 
dwarf galaxy. This scenario has been invoked to explain peculiar properties of
these massive clusters, such as the presence of chemical inhomogeneities, 
the high flattening, and the high central velocity dispersion, among others
\citep{zinnecker+88,hughes+wallerstein00,hilker+richtler00,bekki+freeman03,
bekki+chiba04}. 
While it is not possible to measure the presence or the absence of
metallicity dispersion in HCH99-18,  the three clusters share similar 
properties, with the exception of low ellipticity. The ellipticity 
of HCH99-18 is actually similar to that of M~54 ($\epsilon=0.06$), 
another Galactic globular cluster which is also considered 
to be a former nucleus of a dwarf galaxy. Thus it is possible that HCH99-18 
also formed in a similar way. The 
alternative formation scenario to that of a stripped dwarf galaxy nucleus, 
could be through a merger of two or more young clusters 
\citep{minniti+04_Sersic13,fellhauer+kroupa02}.

This object has the mass comparable to some of the most massive {\it young}
massive clusters in galactic merger remnants, e.g.\ W3 and W30 in NGC~7252 
and G114 in NGC~1613 \citep{maraston+04, bastian+06}. To the best of our 
knowledge it is the brightest and most massive {\it old} globular cluster 
known to date  within the distance of Cen~A, and has similar properties 
to compact massive objects like DGTOs/UCDs observed in the Virgo and Fornax 
clusters, and therefore it definitely warrants further study.

\section{Discussion}
\label{sect:FP}

\begin{figure}
\centering
\resizebox{\hsize}{!}{
\includegraphics[angle=0]{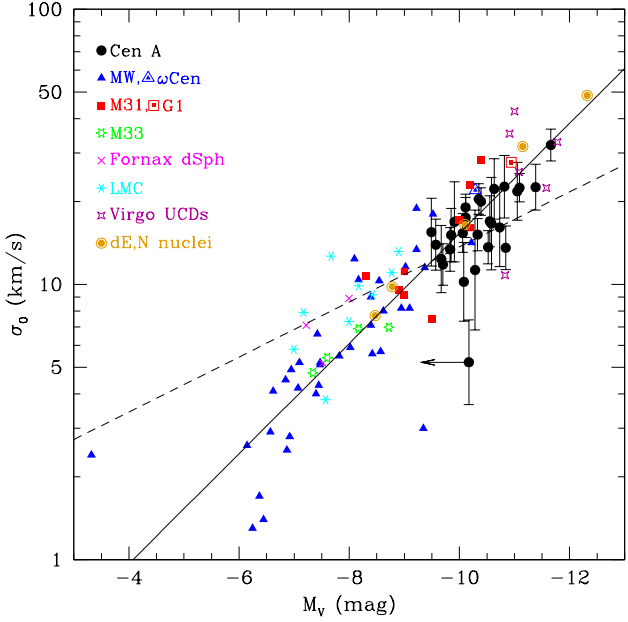}
}
\caption[]{Central velocity dispersion versus absolute $V$ band magnitude 
relation for globular clusters in NGC~5128 (Cen A; black filled circles, this 
work) is compared with old ($age>10$ Gyr) clusters in Local Group galaxies as 
well as with transition objects between globular clusters and dwarf
galaxies, UCDs/DGTOs from the work of \citet{hasegan+05} 
and nuclei of dwarf ellipticals in Virgo \citep{geha+02}. 
The sources of data for the Local Group clusters are as follows: 
the Milky Way, Fornax dSph and LMC clusters are from 
\citet[][]{mclaughlin+vdmarel05},  M31 from \citet[][]{dubath+grillmair97}, 
M33 from \citet[][]{larsen+02}, G1 cluster data are from \citet{meylan+01} 
and $\omega$~Cen from \citet{meylan+95}.
For clarity error-bars are omitted for all but our data.  They 
include the uncertainties in the correction from observed to central $\sigma$. 
The outlying cluster from our sample, R01-122 has its luminosity overestimated 
due to probable blend with foreground stars, as indicated by the arrow.
The dashed line is the Faber-Jackson relation \citep{faber+jackson76} for bright
ellipticals, while the solid line shows the best fit relation for Galactic 
globular clusters from \citet[][]{mclaughlin+vdmarel05}.
{\it (See the electronic edition
of the journal for the colour version of this figure.)}
}
\label{fig:Mv_sigma}
\end{figure}

\begin{figure}
\centering
\resizebox{\hsize}{!}{
\includegraphics[angle=0]{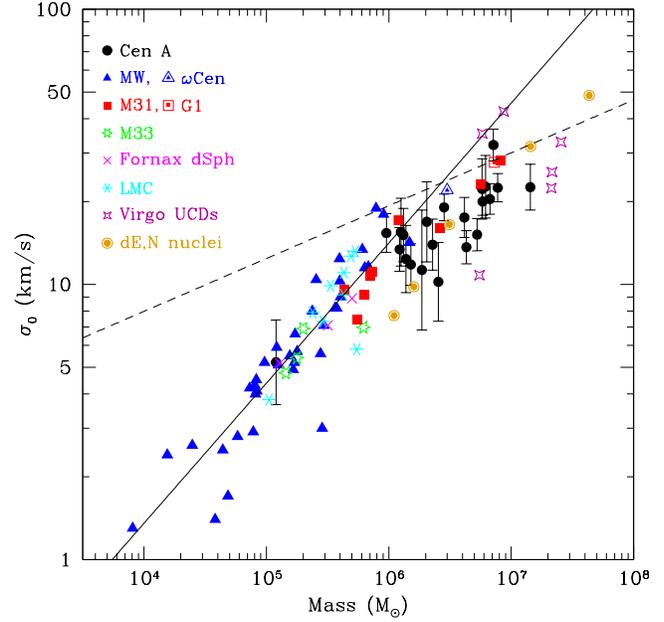}
}
\caption[]{ Central velocity dispersion is plotted as a function of mass. 
For the literature sources for the clusters from the Local Group galaxies and 
UCDs/DGTOs and dE,N nuclei from Virgo clusters (see caption of 
Fig.~\ref{fig:Mv_sigma}). Solid line shows the virial theorem for globular 
clusters and the dashed line is a dependence between mass and $\sigma$ for 
bright elliptical galaxies \citep[Equ.\ 13 and 12 from][]{hasegan+05}.
{\it (See the electronic edition
of the journal for the colour version of this figure.)}
}
\label{fig:Mass_sigma}
\end{figure}

\begin{figure}
\centering
\resizebox{\hsize}{!}{
\includegraphics[angle=0]{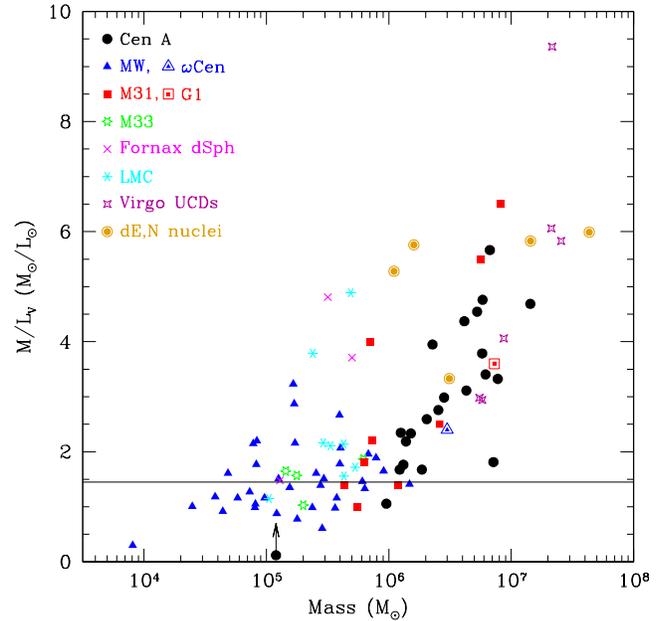}
}
\caption[]{Same as Fig.~\ref{fig:Mass_sigma}, but for mass-to-luminosity as a 
function of mass. Solid line shows the average dynamically determined 
$M/L=1.45$ for a sample of Galactic globular clusters \citep{mclaughlin00}.  
{\it (See the electronic edition
of the journal for the colour version of this figure.)}
}
\label{fig:Mass_ML}
\end{figure}

\begin{figure}
\centering
\resizebox{\hsize}{!}{
\includegraphics[angle=0]{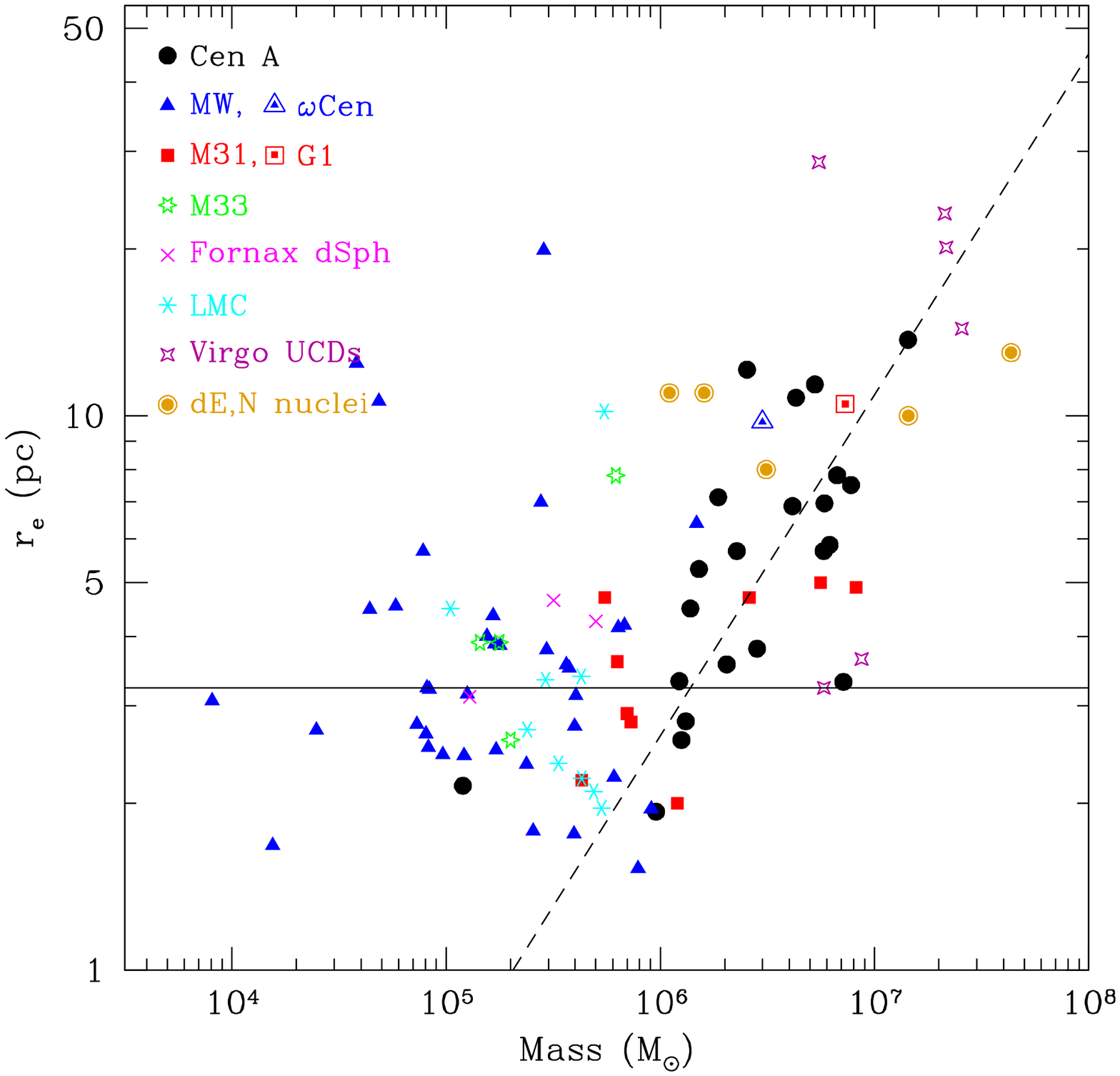}
}
\caption[]{Same as Fig.~\ref{fig:Mass_sigma}, but for effective (half-light) radius 
as a function of mass. Solid line shows the median $r_e=3.2$~pc for 
Galactic globular clusters, 
while the dashed line shows the dependence between 
mass and effective (half-light) radius for bright elliptical galaxies \citep[Equ.\
12 from][]{hasegan+05}.
{\it (See the electronic edition
of the journal for the colour version of this figure.)}
}
\label{fig:Mass_reff}
\end{figure}

Figure~\ref{fig:Mv_sigma} shows the relation between the absolute $V$ magnitude 
and velocity dispersion. The sources of the data compiled from the literature and shown
together with our Cen~A clusters for comparison are given in the caption of the figure.
The dashed line is the Faber-Jackson relation \citep{faber+jackson76} for bright
ellipticals, while the solid line shows the best fit relation for Galactic globular
clusters from \citet{mclaughlin+vdmarel05}. 

The bright globular clusters in NGC~5128 extend the globular cluster 
luminosity-velocity dispersion relation towards brighter objects like DGTOs and 
nucleated dwarf elliptical (dE,N) galaxy nuclei. 

Figures~\ref{fig:Mass_sigma}, \ref{fig:Mass_ML} and \ref{fig:Mass_reff} display the
relations between mass and velocity dispersion, mass-to-luminosity and effective
(half-light) radius, respectively. Our bright clusters in NGC~5128 (Cen~A) are 
plotted together with typical 
old ($age>10$ Gyr) globular clusters from Local Group galaxies: Milky Way, LMC,
Fornax dSph, M31 and M33. For comparison we also plot the transition objects 
between the globular clusters and dwarf galaxies: DGTOs \citep{hasegan+05} and 
nuclei of dE,N \citep{geha+02}. The solid lines are the best 
fit mass-sigma relation for globular clusters, the average dynamically determined 
$M/L=1.45$ \citep{mclaughlin00}, and median $r_e=3.2$~pc, which are for Galactic 
globular clusters independent of mass \citep{mclaughlin+vdmarel05}. 
The dashed lines are the best fit
relations for bright elliptical galaxies as discussed by \citet{hasegan+05}.

In all the above figures (\ref{fig:Mv_sigma}--\ref{fig:Mass_reff}) 
for comparison we also show the locations of the brightest
Milky Way  cluster $\omega$~Cen and M31 cluster G1. 
Since the masses of our clusters were obtained using the virial theorem, we
plot virial masses where available: for $\omega$~Cen, G1, and M33 clusters. 
However, the literature source for the other Milky Way clusters, as well as 
clusters in the LMC, SMC and Fornax \citep{mclaughlin+vdmarel05}, presents only 
masses derived from King model fits, and the same is true for the masses of 
Virgo cluster DGTOs \citep{hasegan+05}. The masses of dE,N galaxy nuclei 
result  from dynamical modeling \citep{geha+02}. When comparing these 
different systems we caution
that \citet{meylan+01} pointed out 
that the masses from King model fits are twice as large 
\citep[however for counter example see][]{larsen+02}.

In Fig.~\ref{fig:Mass_sigma} the departure from the scaling relation for 
globular clusters becomes evident. The brightest clusters in NGC~5128 and 
dE,N nuclei occupy the same part of the diagram, which is shared 
also by some, but not all UCDs/DGTOs. However, as discussed by 
\citet{hasegan+05}  DGTOs from their sample have probably different formation 
mechanisms and thus different properties, with some being more similar to 
typical globular clusters and others either stripped galactic nuclei or 
merged complexes of star clusters.

While the lower mass clusters do not show dependence of the M/L on the 
mass, a quite clear relation emerges for the clusters with masses
larger than $\sim 2 \times 10^6$~M\sun (Fig.~\ref{fig:Mass_ML}).  
This is similar to the nuclei of dwarf galaxies, and UCDs/DGTOs 
\citep{geha+02,hasegan+05} and is also obeyed by the most massive 
clusters in the Milky Way (e.g.\ $\omega$~Cen; framed blue triangle) 
and in M31 (e.g.\ G1; red framed square).

\begin{figure*}
\centering
\resizebox{\hsize}{!}{
\includegraphics[angle=0]{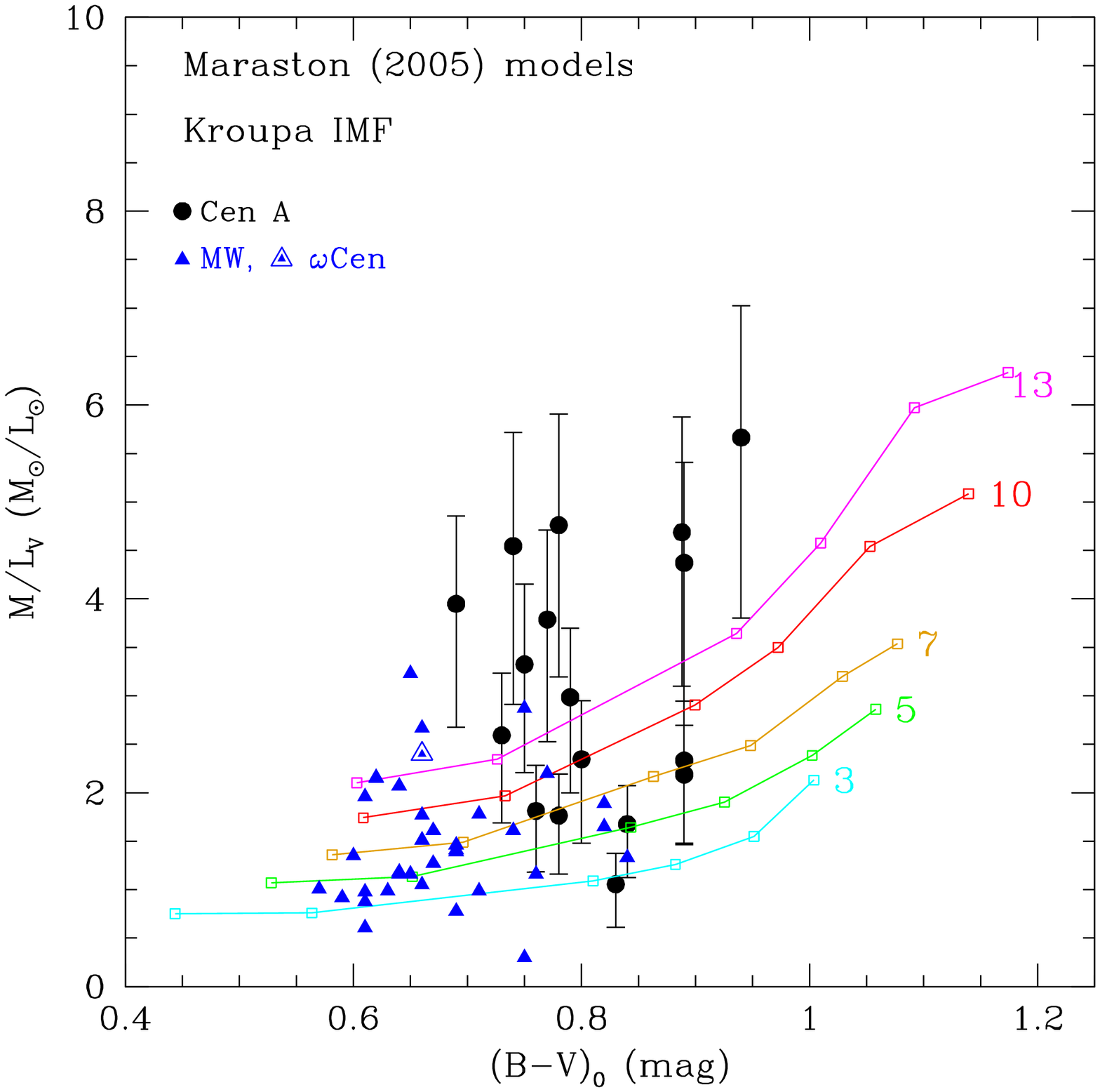}
\includegraphics[angle=0]{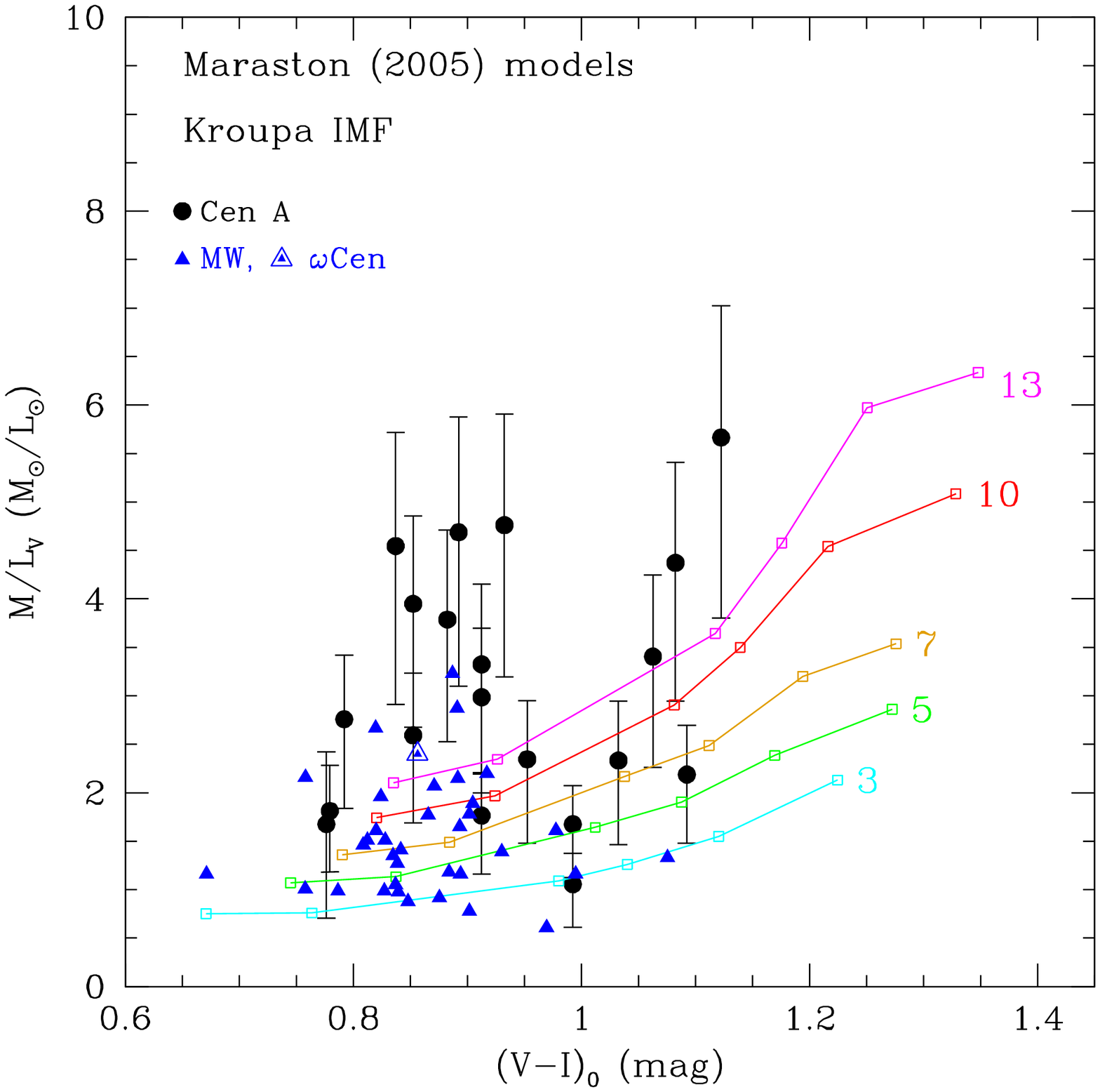}}
\resizebox{\hsize}{!}{
\includegraphics[angle=0]{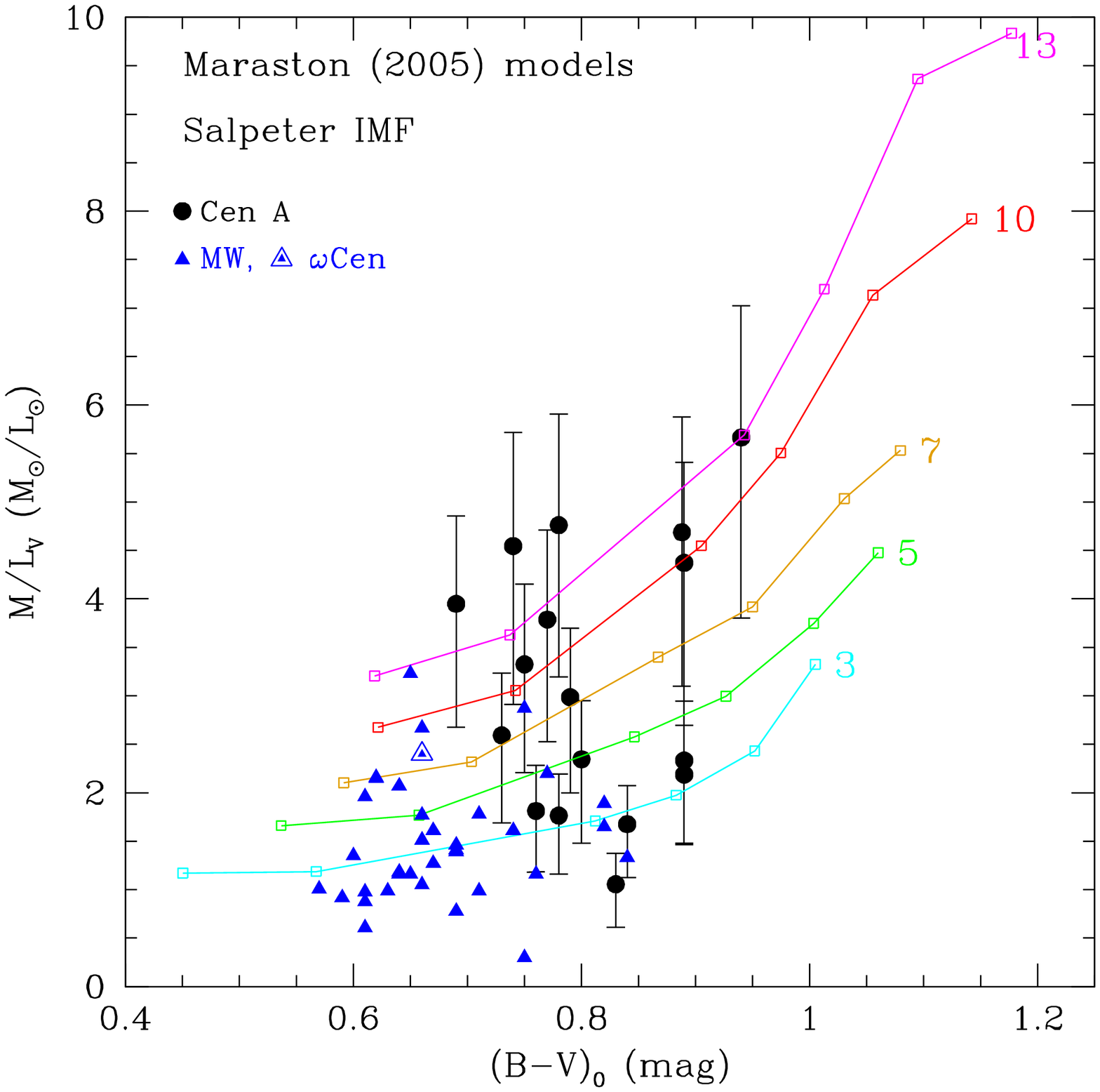}
\includegraphics[angle=0]{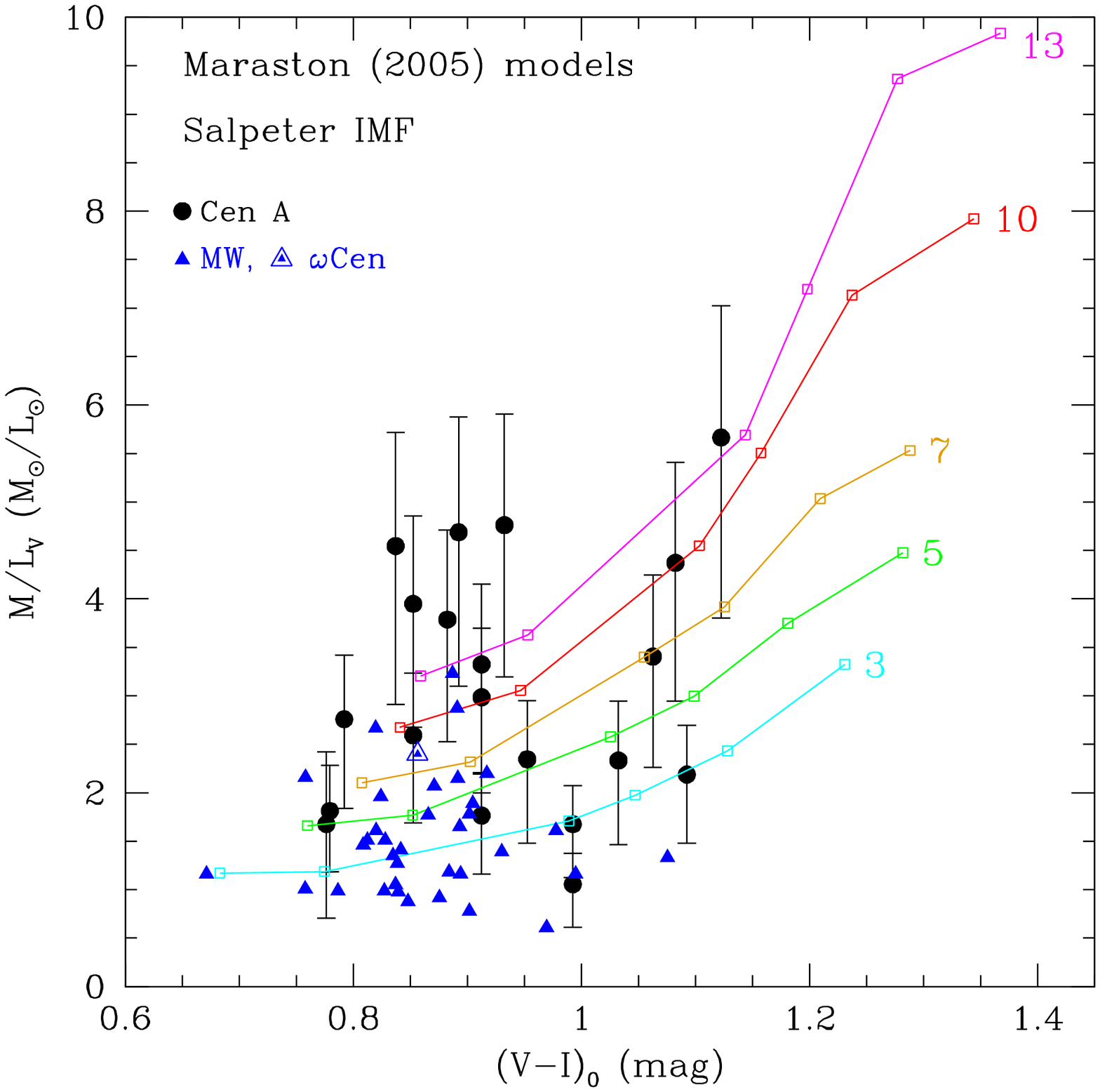}
}
\caption[]{The measured M/L ratios as a function of de-reddened $(B-V)$ (left)
and $(V-I)$ (right) colours for our sample of clusters in Cen~A, and 
old globular clusters in the MW, are compared to 
expected values from SSP models of 
\citet{maraston05} computed for two different IMFs, as shown in each panel. 
The ages of  the theoretical predictions are shown on to the right of the 
curves. The model colours (open squares) from blue to red are for the 
following metallicities:
[Z/H]=$-2.25$, $-1.35$, $-0.33$, $0.0$, $+0.35$ and $0.67$~dex. The colours of
MW clusters are taken from \citet{harris96} 
web\footnote{http://www.physics.mcmaster.ca/~harris/Databases.html} catalogue.
{\it (See the electronic edition
of the journal for the colour version of this figure.)}
}
\label{fig:colour_ML}
\end{figure*}

All Local Group globular clusters plotted in these scaling relation diagrams 
have ages in excess of 10~Gyr.
The ages of Cen~A clusters are not well known. \citet{peng+04GCS} found that the
metal-poor bright Cen~A clusters have ages similar to those of Milky Way clusters, 
while the metal-rich appear younger with ages up to 5 Gyr. Since the $M/L_V$ of a 
population of a given age increases with metallicity, it is in principle 
possible to construct a sample where the more massive globular clusters would 
be more metal-rich, thus having also higher $M/L$ ratios. 
However we note that no relation between the
$(V-I)_0$, nor $(B-V)_0$ colour and $M/L$ is present neither in our data 
(Figure~\ref{fig:colour_ML}) nor was noted by \citet{martini+ho04}. 

Individual spectroscopic metallicities are not available for Cen~A
clusters in the literature. Therefore to explore the age-metallicity
dependence on M/L ratio we use $(B-V)_0$ and $(V-I)_0$ 
colours and compare them to
\citet{maraston05} models in Figure~\ref{fig:colour_ML}. In the upper panels we
plot the models for \citet{kroupa01} initial mass function (IMF), while the 
bottom panels have models calculated with \citet{salpeter55} IMF. Looking 
only at the Cen~A clusters (filled dots with error-bars), the models with 
Salpeter IMF reproduce the range of M/L ratios indicating 
that the clusters with high M/L ratios are old and metal-rich. This result is
very similar to that found for Fornax cluster UCDs by \citet{hilker+07}.
However, we point out that these models predict too high M/L ratios with 
respect to the MW {\it old} globular cluster M/Ls. Models 
with Kroupa IMF pass through the region occupied by MW globulars, but they still
imply too young ages for most of them. \citet{bc03} models with 
\citet{chabrier03} IMF have 
lower M/L ratios and thus fit better the range of M/L values of MW 
globular clusters (e.g.\  see Fig.~11 of \citet{hasegan+05} and Fig.~11 of
\citet{hilker+07}). However, these models do not cover the part of the plane
where Cen A clusters lie. In principle, it is therefore possible by choosing the
IMF and different simple stellar population models to find a good solution for
either Cen~A clusters or MW clusters, but not both.

The average colours of our sample of clusters in NGC~5128 are
$<(B-V)_0>=0.81 \pm 0.07$~mag and $<(V-I)_0>=0.94 \pm 0.10$~mag. This is 
respectively 
0.14 and 0.07~mag redder than the average colours of the MW globulars that we 
use for comparison in Figure~\ref{fig:colour_ML}. Assuming that all the 
clusters have the same old age, the higher average $M/L_V$ ratio for the 
clusters in Cen~A with respect to those in MW can be explained in part with 
their higher metallicity. However, even excluding the reddest clusters from 
our sample, the average $M/L_V$ ratio of the bright clusters in Cen~A is 
still on average higher than that of "normal" globular clusters. As can be 
seen from Figure~\ref{fig:Mass_ML} their M/L ratios cover a range between 
those of "normal" globular clusters and those of UCDs and dE,N nuclei.

We note also that the half-light radius is independent of the mass for the 
low mass clusters, while it increases with the mass for
clusters more massive than $\sim 2 \times 10^6$~M\sun. This shows that, bright, 
massive clusters present a transition type of objects between typical globular 
clusters and more massive DGTOs and dE,N nuclei. The implications of this finding 
for the {\it young} massive clusters has been discussed in detail by 
\citet{kissler-patig+06}. They have argued for the possibility of 
formation of such massive objects through early mergers of low mass stellar 
clusters, which might explain the emergence of mass-radius relation (for 
more details see \citet{kissler-patig+06} and references therein). As an 
alternative speculation Kissler-Patig and collaborators suggest a 
possibility that ``all star clusters form with a primordial mass-radius relation,
but only the most massive clusters are able to retain it against the 
processes that would erase it''. The result presented in 
Figure~\ref{fig:Mass_reff} is consistent with the latter
scenario, which should be further explored theoretically.

Another parameter that might be linked to the formation scenario is ellipticity. 
The high ellipticity of $\omega$~Cen in our Galaxy  and G1 in M31 have been 
frequently mentioned together with other peculiarities shared by these two 
clusters, in the context of the stripped galaxy nucleus formation mechanism 
\citep[e.g.][]{bekki+freeman03,bekki+chiba04}. 
Large fraction of the massive clusters in M31, LMC, and
some in the MW show significant ellipticities 
\citep{geisler+hodge80, harris96, barmby+00}.
Ellipticities of Cen~A clusters have been compared to those of MW globular
clusters and discussed by \citet{holland+99}, \citet{harris+02}, and 
\citet{gomez+06}. They
found a strikingly high fraction of very elongated clusters among the 
luminous clusters in Cen~A. Since our sample contains most of the clusters
already examined by these authors, it is not surprising to reach similar
conclusions. In addition to comparing the ellipticities of clusters as a
function of luminosity we can test whether there is any dependence of
ellipticity on the mass of the cluster - in our sample we find none.

In a recent paper \citet{fellhauer+kroupa06} argue that the tidal heating
during a close passage to the galactic center of a UCD may squeue the 
velocity distribution of its stars and therefore lead to an
overestimation of virial mass and M/L ratio. In these simulations the
more compact and the more massive the object, the smaller is the effect on the
measured velocity dispersion. For the models similar to the UCDs in Virgo and
Fornax clusters, with core radii of the order of 25 pc and masses of
$10^7$~M\sun, only those that pass within 100--1000 pc from the center of 
galaxy can be significantly affected by tidal heating.

In Table~\ref{tab:structpar} we list in 
column 7 the projected galactocentric distance for the clusters in our sample.
No correlation between the mass or M/L ratio and the galactocentric distance is
observed. While without additional simulations it is not clear if and by
how much tidal heating 
could be affecting the velocity dispersion measurements and thus mass
determinations in the relatively (with respect to UCDs) compact globular 
clusters, we cannot exclude a possibility that some of the  
clusters have their masses overestimated due to inadequate assumption of 
virial equilibrium. 
However, given the range of masses, galactocentric distances and the 
relatively compact clusters, the explanation of the mass--$M/L_V$ relation
(Fig.~\ref{fig:Mass_ML}) is unlikely to be due to systematic 
overestimation of mass and $M/L_V$ due to the tidal
heating of the clusters. This relation might instead be connected to
the formation mechanism.


\section{Conclusions}
\label{sect:summary} 

We have presented an analysis of the radial velocities and 
velocity dispersions for 27 bright globular clusters in
the nearby elliptical galaxy NGC~5128. For two targets we have confirmed here 
for the first time their membership in NGC~5128 through 
radial velocity measurements. Also, for 7 clusters we present the first
measurements of their structural parameters from the King profile fitting 
to the high resolution ground based images.

For 22 clusters we combine our new velocity dispersion measurements with 
the information on the structural parameters, either from the literature when 
available or from our own data, and use the virial theorem to derive the 
cluster masses. The masses range from $1.2 \times 10^5$~M\sun, typical of Galactic 
globular clusters, to $1.4 \times 10^7$~M\sun, similar to more massive DGTOs 
and nuclei of dE,N galaxies.

HCH99-18 is the brightest and the most massive cluster in our sample with 
$M_{vir}=1.4\times 10^7$~M\sun. With such a high mass and the $M/L_V$ 
ratio of 4.7 it is a candidate for being 
the remnant nucleus of a stripped dwarf galaxy. The alternative explanation 
could be the merger of two or more young clusters
\citep{minniti+04_Sersic13}. To the best of our knowledge it is the
brightest and most massive {\it old} globular cluster known to date
 within the distance of Cen~A, and it shares similar properties 
with compact massive objects like DGTOs/UCDs observed in the Virgo and Fornax 
clusters. Therefore it definitely warrants further study.

The most striking finding of our study is the emergence of the mass--radius 
and the mass--$M/L_V$ relations for the bright clusters with masses larger 
than $\sim 2 \times 10^6$~M\sun. Figure \ref{fig:Mass_ML} hints to the 
possible existence of two ``populations'' of globular clusters: (1) 
less massive (``normal'') globular clusters, like the ones found 
typically in the Milky Way and M33, with $M/L_V$ roughly independent of 
the mass, and (2) brighter  more massive 
clusters, including our targets from Cen A as well as $\omega$~Cen and G1, 
with $M/L_V$ ratios that seem to increase with increasing
mass. Moreover, population 2 seems to link population 1 with more massive 
objects such as UCDs and dE,N nuclei.
Figure \ref{fig:Mass_reff}, although less clearly, suggests another 
difference between the population 1 with effective radius independent 
of mass, and population 2 with radius increasing with mass. 

This has been previously discussed for the 
{\it young} massive clusters \citep{kissler-patig+06} in galactic mergers 
and for DGTOs in the Virgo cluster  \citep{hasegan+05}. Our results indicate 
that the bright, massive, globular clusters associated
with elliptical galaxies might present the missing link between ``normal'' 
old globular clusters associated with galaxies, {\it young} massive clusters 
formed in mergers and evolved massive objects like UCDs (or DGTOs) associated 
with galaxy clusters.

\begin{acknowledgements}
MR is grateful to Soeren Larsen for helpful hints and discussions and  
acknowledge discussions with the cross-correlation gurus of the 
Geneva Observatory, in particular Stephane Udry and Didier Queloz. 
Many thanks to Andres Jord\'an for interesting
discussions and help with estimating aperture corrections.
MR is also grateful for a stay at the Observatoire de Sauverny supported 
by EPFL. DM is supported by FONDAP Center for Astrophysics 15010003 and by
a Fellowship from the John Simon Guggenheim Foundation. 
We thank an anonymous referee
for the careful reading of the manuscript and suggestions for improvement.

\end{acknowledgements}

\bibliographystyle{aa}
\bibliography{/export/diska/mrejkuba/publications/Article/mybiblio}

\end{document}